# IEEE Copyright Notice

# Non-sliced Optical Arbitrary Waveform Measurement (OAWM) Using a Silicon Photonic Receiver Chip

Daniel Drayss[1,2], Dengyang Fang[1], Christoph Füllner[1],
Wolfgang Freude[1], Sebastian Randel[1], and Christian Koos[1,2]

(1) Institute of Photonics and Quantum Electronics (IPQ) at Karlsruhe Institute of Technology (KIT), 76131 Karlsruhe, Germany
(2) Institute of Microstructure Technology (IMT), Karlsruhe Institute of Technology (KIT), 76344 Eggenstein-Leopoldshafen, Germany
daniel.drayss@kit.edu, christian.koos@kit.edu

*Abstract* — **Comb-based optical arbitrary waveform measurement (OAWM) techniques can overcome the bandwidth limitations of conventional coherent detection schemes and may have disruptive impact on a wide range of scientific and industrial applications. Over the previous years, different OAWM schemes have been demonstrated, showing the performance and the application potential of the concept in laboratory experiments. However, these demonstrations still relied on discrete fiber-optic components or on combinations of discrete coherent receivers with integrated optical slicing filters that require complex tuning procedures to achieve the desired performance. In this paper, we demonstrate the first wavelength-agnostic OAWM front-end that is integrated on a compact silicon photonic chip and that neither requires slicing filters nor active controls. Our OAWM system comprises four IQ receivers, which are accurately calibrated using a femtosecond mode-locked laser and which offer a total acquisition bandwidth of 170 GHz. Using sinusoidal test signals, we measure a signal-to-noise-and-distortion ratio (SINAD) of 30 dB for the reconstructed signal, which corresponds to an effective number of bits (ENOB) of 4.7 bit, where the underlying electronic analog-to-digital converters (ADC) turn out to be the main limitation. The performance of the OAWM system is further demonstrated by receiving 64QAM data signals at symbol rates of up to 100 GBd, achieving constellation signal-to-noise ratios (CSNR) that are on par with those obtained for conventional coherent receivers. In a theoretical scalability analysis, we show that increasing the channel count of non-sliced OAWM systems can improve both the acquisition bandwidth and the signal quality. We believe that our work represents a key step towards out-of-lab use of highly compact OAWM systems that rely on chip-scale integrated optical front-ends.**

*Index Terms*——**Optical sampling, optical receivers, broadband communication, digital signal processing, optical signal processing, photonic integrated circuits, frequency combs, optical arbitrary waveform measurement (OAWM).**

## I. INTRODUCTION

Optical arbitrary waveform measurement (OAWM) based on frequency combs gives access to the full-field information of broadband optical waveforms [1-9]. Applications range from reception of high-speed communication signals [2-9] and elastic optical networking [4] to ultra-broadband photonic-electronic analog-to-digital conversion [10-13] and investigation of ultra-short events in science and technology [1]. Previous demonstrations of comb-based OAWM have relied on spectrally sliced reception, where the broadband optical input signal is first decomposed into a multitude of narrowband spectral slices by appropriate optical filters. These slices are then are individually received by an array of in-phase/quadrature receivers (IQR) using a frequency comb as multi-wavelength local oscillator (LO), and the original waveform is reconstructed by spectral stitching of the received tributaries in the frequency domain [1-5]. However, this concept suffers from the complexity of the underlying high-quality optical filters, which are required both for spectral slicing of the optical signal and for separating the LO comb tones. Specifically, while IQ receivers can be efficiently integrated using readily available high-index-contrast photonic platforms such as silicon photonics (SiP) or indium phosphide (InP), high-quality filters are much more challenging to implement in these material systems due to their sensitivity to fabrication inaccuracies and resulting phase errors. As an example, previous demonstrations of integrated OAWM receivers either relied on InP-based arrayed waveguide gratings (AWG) that required individual phase correction in the various arms [14], or on SiP coupled-resonator optical waveguide (CROW) structures [5,15] that need sophisticated control schemes for thermal tuning. To overcome these challenges, we recently proposed and demonstrated a non-sliced OAWM scheme [6,7], which does not require any high-quality slicing filters. However, while this scheme lends itself to efficient and robust implementation using high-density photonic integrated circuits (PIC), the underlying proof-of-concept experiments still relied on discrete fiber-optic components.

In this paper, we demonstrate the first PIC-based implementation of a non-sliced OAWM front-end [9]. The scheme relies on an array of IQ receivers, which are fed by the full optical waveform and by time-delayed copies of the full LO comb. The electrical signals then contain superimposed mixing products of the various LO tones with the respective adjacent portions of the signal spectrum and allow to reconstruct the full-field information of the incoming waveform using advanced digital signal processing (DSP) [7]. In our work, we demonstrate an integrated OAWM front-end that combines the IQ receiver array with the associated passive components such as power splitters and delay lines on a compact silicon PIC. The front-end does not require any active control of phase shifters and is wavelength-agnostic, thus allowing to receive signals throughout the telecommunication C-band – in sharp contrast to sliced receiver schemes relying on dedicated slicing filters that are either fixed or complex to control. To the best of our knowledge, our experiments represent the first OAWM demonstration with an optical front-end having co-integrated photodetectors. In our proof-of concept experiments, we use four IQ receiver, each relying on photodetectors having a moderate 3 dB bandwidth of less than 20 GHz, to demonstrate an optical acquisition bandwidth of 170 GHz. We analyze the noise and distortions introduced by the OAWM system by measuring an external-cavity laser (ECL) tone tuned to different frequencies, revealing signal-to-noise-and-distortion ratios (SINAD) of approximately 30 dB. This corresponds to an effective number of bits (ENOB) of approximately 4.7 for the overall OAWM system, with the acquisition noise of the underlying analog-to-digital converter (ADC) being the main limitation. The viability of the scheme is shown by reception of various waveforms such as a 100 GBd 64QAM signal or a combination of 60 GBd and 80 GBd 64QAM signals. We finally perform a scalability study, investigating the potential of increased number $N$ of IQR channels and quantify the associated limitations analytically. On the one hand, increasing the channel count offers a path towards efficient bandwidth scaling with linearly increasing hardware. On the other hand, for a fixed overall bandwidth, a higher channel count $N$ allows to relax the bandwidth requirements of a single receiver, such that slower ADC with higher ENOB can be used, thereby increasing the achievable SINAD.

## II. CONCEPT

The concept of the integrated OAWM system is illustrated in Fig. 1 (a). The optical signal under test, $\underline{a}_S(t)$, with spectrum $\tilde{a}_S(f)$ is amplified and fed to the PIC-based OAWM front-end. On the PIC, the signal is split into $N = 4$ copies and routed to an array of integrated IQ receivers (IQR 1...4). An optical frequency-comb generator (FCG) generates $M = 4$ phase-locked optical tones with frequencies $f_\mu$ and with a free spectral range (FSR) $f_{FSR}$ by modulating a continuous-wave tone emitted by a low-linewidth fiber laser. The LO comb is coupled to the PIC, where it is split in $N = 4$ identical copies, which are delayed before being fed to the respective IQR receiver. The individual delays $\tau_\nu$ are approximately evenly distributed over the repetition period of the LO. The in-phase (I) and quadrature (Q) components $I_\nu(t)$ and $Q_\nu(t)$ for IQR $\nu$, $\nu = 1, \dots N$ are extracted from the respective balanced photodiodes and digitized by an array of ADC, Fig. 1 (a).

The mathematical model of non-sliced OAWM is described in more detail in [8] such that we limit our description here to a summary of the essentials. The overall $2N$ recorded baseband spectra $\tilde{I}_\nu(f)$ and $\tilde{Q}_\nu(f)$ for $\nu = 1, \dots, N$ can be related to the overall $M$ frequency down-shifted signal spectra $\tilde{a}_S(f + f_\mu)$ for $\mu = 1, \dots, M$ associated with the LO tones at frequencies $f_\mu$ by a set of $N$ linear equations with frequency-dependent transfer functions $\tilde{\underline{H}}_{\nu\mu}^{(I,U)}(f)$ and $\tilde{\underline{H}}_{\nu\mu}^{(Q,U)}(f)$,

$$\tilde{I}_\nu(f) = \sum_{\mu=1}^{M} \left( \tilde{\underline{H}}_{\nu\mu}^{(I,U)}(f) \tilde{\underline{a}}_S(f + f_\mu) + \tilde{\underline{H}}_{\nu\mu}^{(I,U)*}(-f) \tilde{\underline{a}}_S^*(-f + f_\mu) \right) + \tilde{G}_\nu^{(I)}(f),$$ (1)

$$\tilde{Q}_\nu(f) = \sum_{\mu=1}^{M} \left( \tilde{\underline{H}}_{\nu\mu}^{(Q,U)}(f) \tilde{\underline{a}}_S(f + f_\mu) + \tilde{\underline{H}}_{\nu\mu}^{(Q,U)*}(-f) \tilde{\underline{a}}_S^*(-f + f_\mu) \right) + \tilde{G}_\nu^{(Q)}(f).$$ (2)

In these relations, the frequency-dependent transfer functions $\tilde{\underline{H}}_{\nu\mu}^{(I,U)}(f)$ and $\tilde{\underline{H}}_{\nu\mu}^{(Q,U)}(f)$ comprise the optical characteristics of the respective optical signal path through the setup, the electrical characteristics of the respective IQ receiver IQR $\nu$, as well as the amplitude and the phase and of the associated LO tone at frequency $f_\mu$. The noise added by the receiver system is described by $\tilde{G}_\nu^{(\mathrm{I})}(f)$ and $\tilde{G}_\nu^{(\mathrm{Q})}(f)$ for the in-phase and quadrature component, respectively, and comprises shot noise, thermal noise, noise of electrical amplifiers, and quantization noise. The various noise sources associated with different channels are assumed to be statistically independent. Note that the setup considered here contains an additional optical amplifier, labelled "AMP" in Fig. 1 (a), at the input, which is considered part of the OAWM system. This amplifier adds amplified spontaneous emission (ASE) noise to the incoming signal, which, strictly speaking, needs to be considered for the further analysis of the signal quality. To keep the analysis simple, we assume for now that the optical input signal $\underline{a}_\mathrm{S}(t)$ and the generated LO comb are strong enough to render the ASE noise insignificant. We later conduct a more detailed investigation of the impact of ASE and the associated sensitivity limitations, see Fig. 10 below.

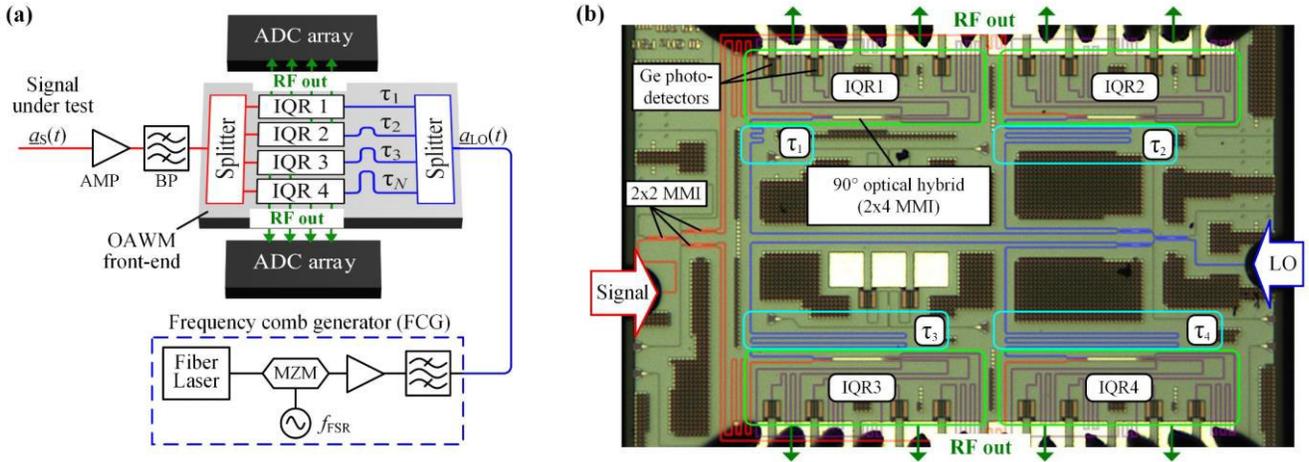

**Fig. 1.** Concept and implementation of our OAWM receiver front-end relying on a silicon photonic integrated circuit (PIC). **(a)** The optical signal under test $\underline{a}_S(t)$ is first amplified and filtered by a bandpass (BP) before being coupled to the OAWM front-end. An optical frequency comb generator (FCG) based on a low-phase-noise fiber laser and a subsequent Mach-Zehnder modulator (MZM) provides four coherent LO tones with free spectral range $f_{FSR}$, which is defined by the modulation frequency. The signal is split into $N = 4$ copies and routed to an array of in-phase quadrature receivers (IQR). The LO is also split, and the copies are delayed by four distinct intervals $\tau_\nu$, $\nu = 1, \dots, 4$, and routed to the IQR array. The radio-frequency (RF) output signals of the IQR are captured by an array of ADC, and the signal under test is reconstructed by digital signal processing (not shown). **(b)** Photograph of the OAWM front-end PIC comprising several 2×2 multi-mode interference couplers (MMI) as power splitters, delay lines for the LO, 90° optical hybrids (2×4 MMI) as well as balanced germanium photodetectors. The RF signals are extracted using RF probes from the top and bottom. The balanced photodetectors (BPD) are biased at −3 V. In our experiments, we use a free spectral range $f_{FSR} = 39.96$ GHz. The IQR are connected to a pair of high-speed real-time oscilloscopes (Keysight UXR series) that serve as ADC.

The relations (1) and (2) can be combined into a single matrix-vector equation. To this end, we interpret the spectra $\tilde{\underline{I}}_\nu(f)$ and $\tilde{\underline{Q}}_\nu(f)$ for $\nu = 1, \ldots, N$ as components of $(N, 1)$ vectors $\tilde{\mathbf{I}}(f)$, $\tilde{\mathbf{Q}}(f)$, respectively, and the transfer functions $\tilde{H}_{\nu\mu}^{(\mathrm{I},\mathrm{t})}(f)$ and $\tilde{H}_{\nu\mu}^{(\mathrm{Q},\mathrm{t})}(f)$ are transformed into two associated $(N, M)$ matrices $\tilde{\underline{\mathbf{H}}}^{(\mathrm{I},\mathrm{t})}(f)$ and $\tilde{\underline{\mathbf{H}}}^{(\mathrm{Q},\mathrm{t})}(f)$. We further define a $(M, 1)$ signal vector $\tilde{\mathbf{A}}_\mathrm{S}(f)$, that comprises all frequency-shifted signal spectra $\tilde{\underline{a}}_\mathrm{S}(f + f_\mu)$, $\tilde{\mathbf{A}}_\mathrm{S}(f) = [\tilde{\underline{a}}_\mathrm{S}(f + f_1) \cdots \tilde{\underline{a}}_\mathrm{S}(f + f_M)]^\mathrm{T}$ [8]. With these definitions, Eq. (1) and (2) can be re-written as

$$\begin{bmatrix} \tilde{\mathbf{I}}(f) \\ \tilde{\mathbf{Q}}(f) \end{bmatrix} = \underbrace{\begin{bmatrix} \tilde{\underline{\mathbf{H}}}^{(\mathrm{I},\mathrm{t})}(f) & \tilde{\underline{\mathbf{H}}}^{(\mathrm{I},\mathrm{t})*}(-f) \\ \tilde{\underline{\mathbf{H}}}^{(\mathrm{Q},\mathrm{t})}(f) & \tilde{\underline{\mathbf{H}}}^{(\mathrm{Q},\mathrm{t})*}(-f) \end{bmatrix}}_{\tilde{\underline{\mathbf{H}}}(f)} \begin{bmatrix} \tilde{\mathbf{A}}_\mathrm{S}(f) \\ \tilde{\mathbf{A}}_\mathrm{S}^*(-f) \end{bmatrix} + \begin{bmatrix} \tilde{\mathbf{G}}^{(\mathrm{I})}(f) \\ \tilde{\mathbf{G}}^{(\mathrm{Q})}(f) \end{bmatrix}. \quad (3)$$

Assuming that all transfer functions are known, $\tilde{\underline{\mathbf{H}}}(f)$ can be inverted for frequencies within the receiver bandwidth $B$, $|f| \le B$, and an estimate $[\tilde{\mathbf{A}}_\mathrm{S}^{(\mathrm{est})}(f)^\mathrm{T}, \tilde{\mathbf{A}}_\mathrm{S}^{(\mathrm{est})*}(-f)^\mathrm{T}]^\mathrm{T}$ for the signal vector $[\tilde{\mathbf{A}}_\mathrm{S}(f)^\mathrm{T}, \tilde{\mathbf{A}}_\mathrm{S}^*(-f)^\mathrm{T}]^\mathrm{T}$ can be reconstructed. This can either be done by calculating the regular matrix inverse $\tilde{\underline{\mathbf{H}}}^{-1}(f)$ of $\tilde{\underline{\mathbf{H}}}(f)$ in case of a square matrix $(N = M)$, or by computing the pseudo-inverse of $\tilde{\underline{\mathbf{H}}}(f)$ in case the number of IQ receivers exceeds that of LO comb tones, $N > M$. The latter leads to a least-square estimate $\tilde{\mathbf{A}}_\mathrm{S}^{(\mathrm{est})}(f)$ of the signal vector from the overdetermined linear system of equations. In the following, we refer to the components $\tilde{\underline{a}}_{\mathrm{S},\mu}^{(\mathrm{est})}(f + f_\mu)$ of the estimated signal vector $\tilde{\mathbf{A}}_\mathrm{S}^{(\mathrm{est})}(f)$ as frequency-shifted signal slices, because they represent the spectral portion around the respective LO tone $f_\mu$ of the original signal spectrum $\tilde{\underline{a}}_\mathrm{S}(f)$. Without loss of generality, we limit the discussion to the case, where the number of IQ receivers $N$ equals the number of LO comb lines $M$, $N = M$, as this configuration leads to the highest acquisition bandwidth, $B_{\mathrm{opt}} \approx M \times f_{\mathrm{FSR}}$ for a given number $N$ of IQ receiver. The inverse of relation (3) can thus be written as,

$$\begin{bmatrix} \tilde{\mathbf{A}}_\mathrm{S}^{(\mathrm{est})}(f) \\ \tilde{\mathbf{A}}_\mathrm{S}^{(\mathrm{est})*}(-f) \end{bmatrix} = \tilde{\underline{\mathbf{H}}}^{-1}(f) \begin{bmatrix} \tilde{\mathbf{I}}(f) + \tilde{\mathbf{G}}^{(\mathrm{I})}(f) \\ \tilde{\mathbf{Q}}(f) + \tilde{\mathbf{G}}^{(\mathrm{Q})}(f) \end{bmatrix} \quad (4)$$

$\underbrace{\phantom{XXXXXXXX}}_{\substack{\text{Reconstructed} \\ \text{signal vector}}}$

Note that Eq. (3) is redundant when evaluated in the full frequency range $f \in [-B, B]$, because $\tilde{\mathbf{I}}(f)$ and $\tilde{\mathbf{Q}}(f)$ are spectra of real-valued signals $I(t)$ and $Q(t)$, such that $\tilde{\mathbf{I}}(f) = \tilde{\mathbf{I}}^*(-f)$ and $\tilde{\mathbf{Q}}(f) = \tilde{\mathbf{Q}}^*(-f)$. Consequently, it is sufficient to evaluate Eq. (4) only for the frequency range $f \in [0, B]$, which yields both $\tilde{\mathbf{A}}_\mathrm{S}^{(\mathrm{est})}(f)$ and $\tilde{\mathbf{A}}_\mathrm{S}^{(\mathrm{est})*}(-f)$ and thus allows to recover the signal vector $\tilde{\mathbf{A}}_\mathrm{S}^{(\mathrm{est})}(f)$ in the full frequency range $f \in [-B, B]$.

Because the transfer functions $\tilde{H}_{\nu\mu}^{(\mathrm{I},\mathrm{t})}(f)$ and $\tilde{H}_{\nu\mu}^{(\mathrm{Q},\mathrm{t})}(f)$ are eventually determined by noisy calibration measurements, the resulting inverse matrix $\tilde{\underline{\mathbf{H}}}^{-1}(f)$ is not known to perfect accuracy. We therefore consider an additional error term $\Delta\tilde{\underline{\mathbf{H}}}^{-1}(f)$ for the signal reconstruction. The estimated signal vector $\tilde{\mathbf{A}}_\mathrm{S}^{(\mathrm{est})}(f)$ is hence not only impaired by noise-related components $\tilde{\mathbf{A}}_\mathrm{G}(f)$ and, but also by crosstalk $\tilde{\mathbf{A}}_\mathrm{X}(f)$ among all spectral slices $\tilde{\underline{a}}_\mathrm{S}(f + f_\mu)$ and conjugate counterparts $\tilde{\underline{a}}_\mathrm{S}(-f + f_\mu)$, see [8],

$$\begin{bmatrix} \tilde{\mathbf{A}}_\mathrm{S}^{(\mathrm{est})}(f) \\ \tilde{\mathbf{A}}_\mathrm{S}^{(\mathrm{est})*}(-f) \end{bmatrix} = \underbrace{\left[ \tilde{\underline{\mathbf{H}}}^{-1}(f) + \Delta\tilde{\underline{\mathbf{H}}}^{-1}(f) \right]}_{} \begin{bmatrix} \tilde{\mathbf{I}}(f) \\ \tilde{\mathbf{Q}}(f) \end{bmatrix}$$

$\underbrace{\phantom{XXXXXXXX}}_{\substack{\text{Reconstructed} \\ \text{signal vector}}}$

$$= \underbrace{\begin{bmatrix} \tilde{\mathbf{A}}_\mathrm{S}(f) \\ \tilde{\mathbf{A}}_\mathrm{S}^*(-f) \end{bmatrix}}_{\text{Signal vector}} + \underbrace{\Delta\tilde{\underline{\mathbf{H}}}^{-1}(f)\tilde{\mathbf{H}}(f)\begin{bmatrix} \tilde{\mathbf{A}}_\mathrm{S}(f) \\ \tilde{\mathbf{A}}_\mathrm{S}^*(-f) \end{bmatrix}}_{\text{Crosstalk:} \left[ \tilde{\mathbf{A}}_\mathrm{X}(f)^\mathrm{T} \quad \tilde{\mathbf{A}}_\mathrm{X}^*(-f)^\mathrm{T} \right]^\mathrm{T}}$$

$$+ \underbrace{\left[ \tilde{\underline{\mathbf{H}}}^{-1}(f) + \Delta\tilde{\underline{\mathbf{H}}}^{-1}(f) \right]\begin{bmatrix} \tilde{\mathbf{G}}^{(\mathrm{I})}(f) \\ \tilde{\mathbf{G}}^{(\mathrm{Q})}(f) \end{bmatrix}}_{\text{Noise:} \left[ \tilde{\mathbf{A}}_\mathrm{G}(f)^\mathrm{T} \quad \tilde{\mathbf{A}}_\mathrm{G}^*(-f)^\mathrm{T} \right]^\mathrm{T}}.$$

$$(5)$$

After reconstructing the frequency-shifted spectral slices $\tilde{\underline{a}}_{\mathrm{S},\mu}^{(\mathrm{est})}(f + f_\mu)$, that are the components of the reconstructed signal vector $\tilde{\mathbf{A}}_\mathrm{S}^{(\mathrm{est})}(f)$ in the range $f \in [-B, B]$, we undo the frequency shift numerically and stitch the resulting spectra by performing a weighted average to obtain an estimate $\tilde{\underline{a}}_\mathrm{S}^{(\mathrm{est})}(f)$ for the input spectrum, see [8] for a more detailed description. The time-domain waveform $\underline{a}_\mathrm{S}(t)$ is then recovered by an inverse Fourier transform.

For a practical implementation, we need to determine the transfer matrix $\tilde{\underline{\mathbf{H}}}(f)$ that is composed of the various transfer functions $\tilde{H}_{\nu\mu}^{(\mathrm{I},\mathrm{t})}(f)$ and $\tilde{H}_{\nu\mu}^{(\mathrm{Q},\mathrm{t})}(f)$. Importantly, the transfer functions are impacted by slow drifts of the optical phases $\varphi_{\mathrm{F},\nu}^{(\mathrm{t})}$ in the various receiver paths, $\nu = 1, \ldots, N$, as well as by unknown initial phases $\varphi_{\mathrm{LO},\mu}^{(\mathrm{t})}$ of the various LO comb lines at frequencies $f_\mu$ $\mu = 1, \ldots, M$, and we have to correct for these impairments during signal reconstruction. To this end, we separate the transfer functions $\tilde{H}_{\nu\mu}^{(\mathrm{I},\mathrm{t})}(f)$, $\tilde{H}_{\nu\mu}^{(\mathrm{Q},\mathrm{t})}(f)$ associated with the various receivers IQR$_\nu$ $\nu = 1, \ldots, N$ and comb lines $f_\mu$, $\mu = 1, \ldots, M$, in a frequency- dependent and time-invariant component $\tilde{H}_{\nu\mu}^{(\mathrm{I})}(f)$, $\tilde{H}_{\nu\mu}^{(\mathrm{Q})}(f)$, that is determined in a separate calibration measurement, see Sect. IV below, and in two slowly time-variant factors $H_{\mathrm{F},\nu}^{(\mathrm{t})} = \exp(\mathrm{j}\varphi_{\mathrm{F},\nu}^{(\mathrm{t})})$ and $H_{\mathrm{LO},\mu}^{(\mathrm{t})} = \exp(\mathrm{j}\varphi_{\mathrm{LO},\mu})$, that describe the phase drift accumulated along the various detection paths as well as the

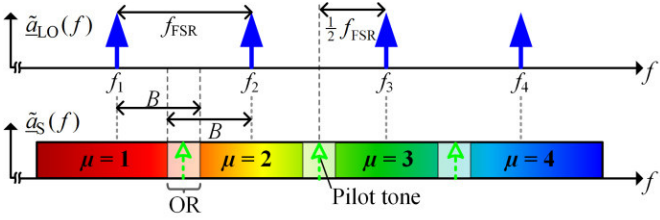

**Fig. 2**. Generation of partially redundant signal components by down-conversion of a broadband optical input signal with spectrum $\underline{\tilde{a}}_{S}(f)$ using an LO comb with spectrum $\underline{\tilde{a}}_{LO}(f)$ and ADC with bandwidth $B > f_{FSR}/2$. Within the overlap regions (OR), $f_{OR} \in \left[ f_{\mu+1} - B, f_{\mu} + B \right]$, $\mu = 1, 2, 3$, spectral components of the optical input signal $\underline{\tilde{a}}_{S}(f)$ are down-converted to the baseband twice, since the mixing products $\left| f_{OR} - f_{\mu} \right|$ and $\left| f_{OR} - f_{\mu+1} \right|$ with both adjacent LO tones at frequencies $f_{\mu}$ and $f_{\mu+1}$ fall into the detection bandwidth $B$ of the corresponding IQ receiver. In case the optical input signal does not comprise spectral components within the OR, we may choose to add pilot tones (dashed green arrows) to ensure that there exists redundancy in the downconverter baseband signals $\underline{\hat{I}}_{\nu}(f)$ and $\underline{\hat{Q}}_{\nu}(f)$ as needed for the phase drift compensation.

initial phase of each LO comb line, respectively [7]. The transfer functions can thus be written as

$$
\begin{aligned}
\underline{\tilde{H}}_{\nu\mu}^{(I,t)}(f) &= \underline{H}_{F,\nu}^{(t)} \times \underline{H}_{LO,\mu}^{(t)} \times \underline{\tilde{H}}_{\nu\mu}^{(I)}(f) = e^{j\left( \varphi_{F,\nu}^{(t)} + \varphi_{LO,\mu} \right)} \underline{\tilde{H}}_{\nu\mu}^{(I)}(f), \\
\underline{\tilde{H}}_{\nu\mu}^{(Q,t)}(f) &= \underline{H}_{F,\nu}^{(t)} \times \underline{H}_{LO,\mu}^{(t)} \times \underline{\tilde{H}}_{\nu\mu}^{(Q)}(f) = e^{j\left( \varphi_{F,\nu}^{(t)} + \varphi_{LO,\mu} \right)} \underline{\tilde{H}}_{\nu\mu}^{(Q)}(f),
\end{aligned}
\tag{6}
$$

Where the complex-valued factors $\underline{H}_{F,\nu}^{(t)}$ and $\underline{H}_{LO,\nu}^{(t)}$ are either fixed or drift very slowly with time, see Fig. 3(d) below, and can therefore be considered constant during one recording with a typical length a few microseconds. We further assume that the LO comb tones are phase locked and do not drift independently such that the corresponding time-domain pulse shape is stable. In this case, the only free LO parameter is the relative temporal position $\tau_{LO}$ of the pulse train within the recording acquired by the ADC array. Consequently, we may reduce the number of free parameters by setting $\varphi_{LO,\mu} = 2\pi f_{FSR} \tau_{LO} \mu$. To estimate and compensate for these drifts, we exploit redundant information that is comprised in the baseband signals $I(t)$ and $Q(t)$, if the ADC bandwidth $B$ exceeds $f_{FSR}/2$. In this case, spectral components of the optical input signal that are located in so-called overlap regions (OR), $f_{OR} \in \left[ f_{\mu+1} - B, f_{\mu} + B \right]$, are down-converted to the baseband twice since the mixing products $\left| f_{OR} - f_{\mu} \right|$ and $\left| f_{OR} - f_{\mu+1} \right|$ with both adjacent LO tones at frequencies $f_{\mu}$ and $f_{\mu+1}$ fall into the detection bandwidth $B$ of the corresponding IQ receiver, see Fig. 2. This creates signal components with redundant information, which allows us to quantify and to compensate the phase drift along the various detection paths as well as the temporal position $\tau_{LO}$ of the pulse train within the recording [8]. Note that the presented non-sliced OAWM scheme is closely related to asynchronous time interleaving that is used in high-speed digital oscilloscopes [16]. However, for the optical implementation used here, the compensation of phase drifts in the various detection paths is crucial, especially when measuring unknown arbitrary waveforms that do not allow to use phase correction algorithms that are available for data signals only [17-19].

## III. INTEGRATED OAWM RECEIVER FRONT-END

The OAWM receiver concept illustrated in Fig. 1 (a) has been implemented using the silicon photonic integration platform, see Fig. 1 (b) for a photograph of the associated PIC. The PIC comprises grating couplers for feeding in the signal (left) and the LO (right). Alternatively, signal and LO can be launched via edge couplers, which are prepared for future optical packaging with photonic wire bonds (PWB) [20]. The power splitters rely on 2×2 multi-mode interference (MMI) couplers, and the 90° optical hybrids exploit 2×4 MMI that establish the desired 90° phase relationship between its paired outputs for the in-phase $I_{\nu}(t)$ and quadrature $Q_{\nu}(t)$ signals, such that no active phase shifters are required. In the context of our calibration measurements with a known reference waveform, we measure the IQ-phase for all IQR to be in the range of 84° to 89°. Each balanced photodetector (BPD) consists of two Germanium photodiodes that are reverse biased at −3 V. The read-out pads of all BPD are contacted using a pair of 4×GSG (ground signal ground) probes, which are connected via 70 cm-long coaxial cables to two synchronized oscilloscopes (Keysight UXR-series) serving as ADC. The digital data is processed offline in Matlab. In all our measurements we use an LO comb with an FSR of $f_{FSR} = 39.96$ GHz .

## IV. CALIBRATION

For accurate signal reconstruction, the transfer functions $\underline{\tilde{H}}_{\nu\mu}^{(I)}(f)$ and $\underline{\tilde{H}}_{\nu\mu}^{(Q)}(f)$ must be determined. To this end, we rely on a known optical reference waveform (ORW), which is generated by a femtosecond laser (Menhir 1550) with a repetition rate of $f_{ORW} = 250$ MHz and which allows us to spectrally sample the transfer function at discrete points spaced by $f_{ORW}$. The reference waveform has been characterized independently by a frequency-resolved optical gating (FROG) measurement. Compared to our first demonstration of the non-sliced OAWM scheme [7], we improve our calibration technique in terms of SNR and linearity by dispersing the ORW using a 20 km single mode fiber (SMF) before feeding it to the OAWM system. The chirped optical pulses have a significantly reduced peak power compared to unchirped pulses such that saturation effects of the photodetectors can be avoided without reducing the average signal power. At the same time, the peak-to-average power ratio (PAPR) of the generated photocurrents is reduced. This increases the SNR of the digitized waveforms, because the quantization noise scales in proportion to the full-scale voltage $U_{FS}$, which needs to be adjusted according to the peak amplitude. However, the so measured transfer functions still include the quadratic phase profile of the 20 km fiber. We therefore characterize the phase profile imprinted by the SMF by using two fibers, SMF1 and SMF2, with approximately 10 km length each. We perform a calibration with only SMF1, only SMF2, and SMF1 and SMF2 concatenated after the ORW source and then extract the system's phase response $\varphi_{\nu\mu}^{(I)}(f) = \arg\left( \underline{\tilde{H}}_{\nu\mu}^{(I)}(f) \right)$ and $\varphi_{\nu\mu}^{(Q)}(f) = \arg\left( \underline{\tilde{H}}_{\nu\mu}^{(Q)}(f) \right)$ by combing the three independent measurements,

$$\varphi_{\nu\mu}^{(\mathrm{I})}(f) = \underbrace{\left(\varphi_{\nu\mu}^{(\mathrm{I})}(f) + \varphi_{\mathrm{SMF1}}(f)\right)}_{\text{Measured with SMF1 (10km)}} + \underbrace{\left(\varphi_{\nu\mu}^{(\mathrm{I})}(f) + \varphi_{\mathrm{SMF2}}(f)\right)}_{\text{Measured with SMF2 (10km)}}$$
$$- \underbrace{\left(\varphi_{\nu\mu}^{(\mathrm{I})}(f) + \varphi_{\mathrm{SMF1}}(f) + \varphi_{\mathrm{SMF2}}(f)\right)}_{\text{Measured with SMF1 and SMF2 (20km)}}. \tag{7}$$

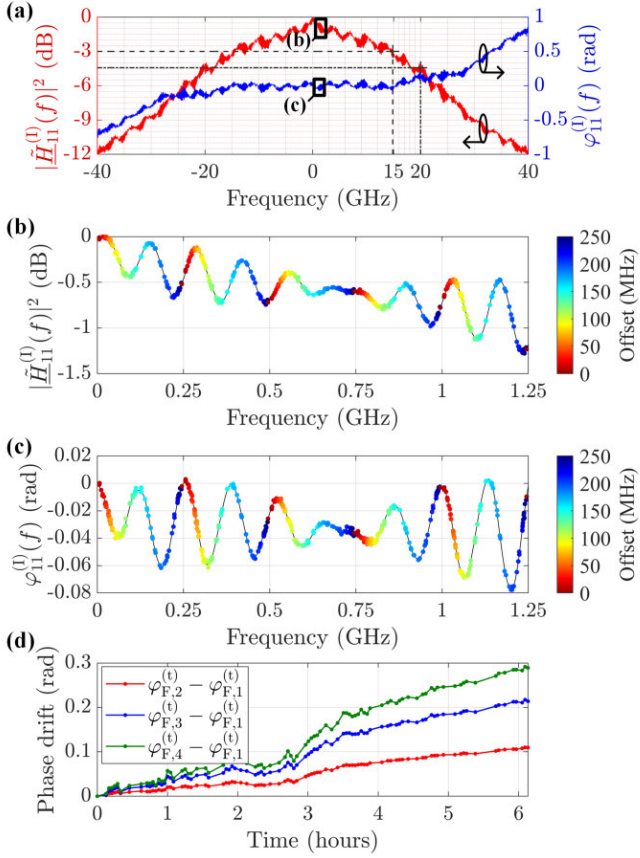

**Fig. 3.** Calibration measurement using a known optical reference waveform (ORW) for extracting the frequency-dependent transfer functions $\tilde{H}_{\nu\mu}^{(\mathrm{I})}(f)$ and $\tilde{H}_{\nu\mu}^{(\mathrm{Q})}(f)$ associated with the various detection paths (subscript $\nu$) and the various LO tones (subscript $\mu$). **(a)** Power transfer function $\left|\tilde{H}_{11}^{(\mathrm{I})}(f)\right|^2$ (red) and phase response $\varphi_{11}^{(\mathrm{I})}(f)$ (blue) as extracted from the measured transfer function $\tilde{H}_{11}^{(\mathrm{I})}(f) = \left|\tilde{H}_{11}^{(\mathrm{I})}(f)\right|\exp\!\left(\mathrm{j}\varphi_{11}^{(\mathrm{I})}(f)\right)$ associated with the detection path leading to first IQ receiver and the first comb tone. We observe a 3 dB bandwidth of approximately 15 GHz and a roll-off of 4.5 dB at $f_{\mathrm{FSR}}/2 = 20\,\mathrm{GHz}$. The relatively small bandwidth of the photodetectors and the pronounced ripples on top of the amplitude and phase responses can be attributed to reflections in the probe and the 70 cm-long RF cables, originating from poor matching of the high-impedance photodetectors to the 50 Ω input of the oscilloscope. Improved electronic read-out could provide the full RF bandwidth of the silicon photonic photodetectors, that is specified to 40 GHz by the foundry. **(b)** Zoom-in of the power transfer function from 0 to 1.25 GHz, see black box in Subfigure (a). The color-coded dots represent individual calibrations that are recorded with different frequency offsets between ORW and LO. The individual calibrations are combined in a post-processing step that compensates for the time-variant phase drifts. As a result, a calibration with high SNR and high spectral resolution is obtained. **(c)** Zoom-in of the phase transfer function from 0 to 1.25 GHz, corresponding again to the black box in (a). **(d)** Relative drift $\varphi_{\mathrm{F},\nu}^{(\mathrm{t})} - \varphi_{\mathrm{F},1}^{(\mathrm{t})}$ of the optical phase parameters $\varphi_{\mathrm{F},\nu}^{(\mathrm{t})}$ over several hours. The phase drifts increases approximately in proportion to the length differences of the respective delay lines, see Fig. 1 (b) for delays $\tau_1$ to $\tau_N$. The observed phase drift is mainly induced by thermal effects and could be significantly reduced by actively stabilizing the chip temperature.

In Fig. 3 (a), we exemplarily show the measured transfer functions $\tilde{H}_{11}^{(\mathrm{I})}(f) = \left|\tilde{H}_{11}^{(\mathrm{I})}(f)\right|\exp\!\left(\mathrm{j}\varphi_{11}^{(\mathrm{I})}(f)\right)$ for the in-phase component $I_1$ associated with the first IQ receiver after removing the phase transfer function of the dispersive fibers. The power response $\left|\tilde{H}_{11}^{(\mathrm{I})}(f)\right|^2$, red trace in Fig. 3 (a), comprises the roll-off of all components (BPD, probe, 70 cm-long RF cable, coaxial adaptors), and reaches $-4.5\,\mathrm{dB}$ at the minimum required ADC bandwidth $B_{\mathrm{RF}} \geq f_{\mathrm{FSR}}/2 \approx 20\,\mathrm{GHz}$. The phase response $\varphi_{11}^{(\mathrm{I})}(f)$, blue trace in Fig. 3 (a), is approximately flat within the bandwidth of interest. The measured bandwidth is presumably limited by reflections inside the probe, originating from an impedance mismatch of the high-impedance photodetectors and the 50 Ω transmission line.

To obtain a calibration with sufficient spectral resolution, we resort to a multi-shot calibration technique, i.e., we acquire several calibrations while varying the frequency offset between the ORW and the LO. This is illustrated in zoom-in shown in Fig. 3 (b), where the color-coding indicates the frequency offset between the ORW and the LO at which the respective data point was taken. The region covered by the 1.25 GHz-wide zoom-in is indicated by a box in Fig. 2 (a), and the associated phase is sown in Fig. 3 (c). Note that the fine ripples on top of the amplitude and phase response in Fig. 3(b) and (c) can be attributed to reflections in the 70 cm-long RF cables connecting the high-impedance photodetectors via the RF probe to the 50 Ω input of the oscilloscope.

Figure 3 (d) shows the slow drift of the optical phase parameters $\varphi_{\mathrm{F},\nu}^{(\mathrm{t})} = \arg(\underline{H}_{\mathrm{F},\nu}^{(\mathrm{t})})$ relative to the phase $\varphi_{\mathrm{F},1}^{(\mathrm{t})}$ associated with the detection path of the first IQ receiver, recorded over a period of six hours. The digital signal reconstruction algorithm estimates these phase drifts for each recording individually without requiring an additional calibration measurement, see [8]. Because all IQR are integrated on a single PIC, the observed relative phase drift $\varphi_{\mathrm{F},\nu}^{(\mathrm{t})} - \varphi_{\mathrm{F},1}^{(\mathrm{t})}$ in Fig. 3 (d) is rather small. This allows the DSP to average the phase information obtained from subsequent recordings acquired within a few seconds, thereby making the final phase estimate more robust.

## V. EXPERIMENTAL RESULTS

We test the OAWM system using different optical data signals that were generated by high-speed IQ modulators and electrical arbitrary-waveform generators (Keysight M8194A). Figure 4 (a) shows the power spectrum of a reconstructed 100 GBd 64QAM signal along with the corresponding constellation diagram, from which we estimate a constellation SNR (CSNR) of 19.3 dB. The CSNR is the square of the reciprocal of the error vector magnitude (EVM) normalized to the average signal power, $\mathrm{CSNR} = 10 \times \log_{10}\left(1/\mathrm{EVM}_{\mathrm{m}}^2\right)$ [21]. As a reference, we measure the same signal using a single intradyne IQ receiver based on discrete high-end 100 GHz photodiodes, obtaining a CSNR of 18.8 dB, which is 0.5 dB lower than the value obtained with for the non-sliced OAWM system. This emphasizes the ability of our OAWM system to offer a good signal quality over large detection bandwidths, even though the individual photodiodes were limited in bandwidth to approximately 15 GHz, see Fig. 3 (a). These findings are well in line with previous results that relied on the same transmitter setup, showing a CSNR of 18.6 dB for a single IQ receiver and CSNR of 19.2 dB for a four-channel spectrally sliced OAWM system [5]. We additionally record the acquisition noise $\underline{\tilde{\mathbf{G}}}_{\text{acq}}(f)$ of the ADC (Keysight UXR series oscilloscope) with all optical signals being disconnected. As the shot-noise is significantly below the acquisition noise, we can approximate the overall receiver noise $\underline{\tilde{\mathbf{G}}}(f)$ by the recorded acquisition noise $\underline{\tilde{\mathbf{G}}}_{\text{acq}}(f)$, $\underline{\tilde{\mathbf{G}}}(f) \approx \underline{\tilde{\mathbf{G}}}_{\text{acq}}(f)$. After applying the same DSP to the noise recordings as previously to the actual signal recordings, we obtain the stitched acquisition noise $\tilde{a}_{\text{G}}(f)$, gray trace in Fig. 4 (a). The noise floor has a periodic frequency dependence, originating from the spectral stitching of the various slices and from the roll-off of the IQ receivers, which is digitally compensated for each slice. In Fig. 4 (b), we show another example, where we record an 80 GBd and a 60 GBd 64QAM signal simultaneously within the optical acquisition bandwidth of 170 GHz. Due to the lower bandwidths of the individual data signals, the quality of the transmitted signals is improved, and the CSNR obtained for the OAWM system increases to 21.4 dB and 22.2 dB, respectively. The receiver bandwidth can well compete with previous demonstrations of integrated spectrally sliced OAWM receivers that still relied on external photodiodes [5].

The data signals shown in Fig. 4 include noise and distortions from the signal generator, which limits the maximal achievable CSNR. To characterize the noise and distortions solely introduced by the non-sliced OAWM system, we use a tunable external cavity laser with a large carrier-to-noise ratio (OCNR) as a monochromatic optical signal source. We can tune the emission frequency of the laser to characterize the performance of the OAWM system at different input frequencies. We adjust the vertical scale of the oscilloscope based on the RF signal with the largest amplitude so that no clipping occurs. As a result, the quantized signals fill between 55% and 100% of the oscilloscope's full range, depending on the frequency of the signal laser. As the narrowband laser signal does not have any spectral components within the overlap region that is exploited for the estimation of the time variant factors $\underline{H}_{\text{F},\nu}^{(\text{t})}$ and $\underline{H}_{\text{LO},\mu}^{(\text{t})}$ in Eq. (6), we add low-power pilot tones (see green dashed arrows in Fig. 2), each ~43 dB lower than the signal.

Fig. 5(a) shows a reconstructed spectrum of an exemplary measurement, where the monochromatic signal (red), distortions (green, magenta, blue, yellow) and noise (gray) are highlighted in different colors. The vertical axis is the spectral power, normalized to the signal, and the horizontal axis gives the frequency offset from the lower edge $f_{\text{ref}} \approx 192.38$ THz of the acquisition range. We separate the following distortions:

1. The four pilot tones (green). Their total power is 37.3 dB lower than the power of the signal (red). Note that we added four tones, even though only three overlap regions (OR) exist. The fourth tone close to the upper edge of the acquisition range is not required for phase-drift compensation but leads to a better convergence of the phase estimation algorithm.

2. The calibration crosstalk $\underline{\tilde{\mathbf{A}}}_{\text{X}}(f)$, see Eq. (5), can be separated into two contributions: Crosstalk from different spectral slices $\underline{\tilde{a}}_{\text{S}}(f + f_{\mu})$ (magenta), which is 45.9 dB below the signal, and IQ crosstalk from the associated mirrored complex-conjugate components $\underline{\tilde{a}}_{\text{S},\mu}^*(-f + f_{\mu})$ (cyan), which is 42.7 dB below the signal. These crosstalk contributions are a consequence of an inaccurate calibration of $\underline{H}_{\nu\mu}^{(\text{I})}(f)$ and $\underline{H}_{\nu\mu}^{(\text{Q})}(f)$, and an inaccurate parameter estimation, $\underline{H}_{\text{F},\nu}^{(\text{t})}$, leading to $\Delta\underline{\tilde{\mathbf{H}}}^{-1}(f) \neq 0$ in Eq. (5).

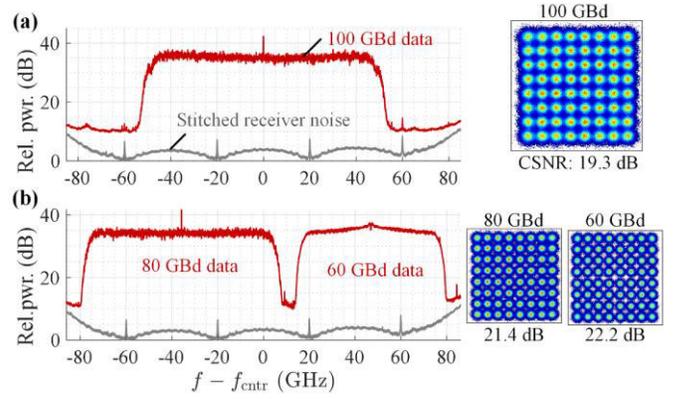

**Fig. 4.** Acquisition of broadband data signals. **(a)** Power spectrum (resolution bandwidth 100 MHz) of a reconstructed 100 GBd 64QAM data signal (red) along with the stitched receiver noise $\tilde{a}_{\text{G}}(f)$ (gray) and the corresponding constellation diagrams. The constellation-signal-to-noise ratio (CSNR) amounts to 19.3 dB. The horizontal axis gives the frequency offset between the optical frequency $f$ and the center frequency $f_{\text{cntr}} \approx 192.47$ THz. **(b)** Power spectrum of a reconstructed signal (red) comprising two data streams (80 GBd 64QAM and 60 GBd 64QAM) along with the stitched receiver noise $\tilde{a}_{\text{G}}(f)$ (gray) and the corresponding constellation diagrams.

3. The clock of the back-end ADC and its mixing products with the signal (yellow), −49 dB.

All distortions add up to a total power which is 35.5 dB below the signal and 3.9 dB below the total power associated with the stitched receiver noise $\tilde{a}_{\text{G}}(f)$, gray in Fig. 5 (a). The signal-to-noise-and-distortion ratio (SINAD) is calculated by dividing the signal power $P_{\text{signal}}$ by the power of the distortions $P_{\text{distortions}}$ and the noise $P_{\text{noise}}$,

$$\text{SINAD}_{\text{dB}} = 10 \log_{10}\left(\frac{P_{\text{signal}}}{P_{\text{noise}} + P_{\text{distortions}}}\right). \tag{8}$$

The SINAD amounts to 30.1 dB, which corresponds to $\text{ENOB} = \left(\text{SINAD}_{\text{dB}} - 1.76\,\text{dB}\right)/6.02\,\text{dB} = 4.7$ bit [40]. Using the same procedure, we characterize our system for different optical frequencies of the monochromatic test signal, see Fig. 5 (b). The color code for the individual distortions remains unchanged, and one may still refer to Fig. 5 (a) for the legend. The solid lines represent the average result obtained from two measurements performed at approximately the same input frequency. We additionally plot the quantity -SINAD$_{\text{dB}}$ as a function of the input frequency in blue in Fig. 5 (b). The average SINAD over all input frequencies is 29.7 dB. Note that the low crosstalk (cyan, magenta) validates the linear system model according to Eqs. (1) and (2), the calibration procedure, and the phase-compensation technique.

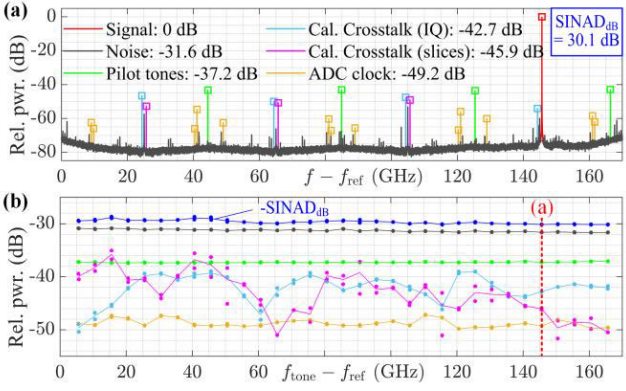

**Fig. 5.** Characterization of the OAWM system by measuring single continuous-wave (CW) laser tones. **(a)** Exemplary spectrum of reconstructed CW signal at 145.6 GHz. The spectrum is normalized to the signal peak (red), and the horizontal axis gives the frequency offset from a reference frequency $f_{ref} \approx 192.38$ THz that was chosen to be the lower edge of the acquisition range. Different distortions are marked. Green: Optical pilot tones added to ease estimation of $\underline{H}_{F,\nu}^{(0)}$ and $\underline{H}_{LO,\mu}^{(0)}$. Note that the rightmost tone at approximately 166.3 GHz is not required for phase-drift compensation but leads to a better convergence of the phase estimation algorithm. Magenta and cyan: Crosstalk due to an imperfect calibration and parameter estimation. Yellow: ADC clock tones after signal reconstruction. Gray: Stitched receiver noise $\underline{\tilde{a}}_{G}(f)$. **(b)** Power of noise and different distortions as a function of the frequency offset $f_{tone} - f_{ref}$, where $f_{tone}$ is the optical frequency of the single-tone laser signal. The color code for the individual distortions remains unchanged, see Subfigure (a) for the legend. The sum of relative noise and all distortions results in the blue trace (-SINAD$_{dB}$).

## VI. Scalability Study

The motivation for increasing the channel count $N$ of parallel IQ receivers in an OAWM system can be twofold: On the one hand, more IQ receivers offer a larger total optical acquisition bandwidth $B_{opt} \approx N \times 2B$, which cannot be achieved by using a single-channel IQ receiver. On the other hand, a given optical acquisition bandwidth $B_{opt}$ can be measured with a better SNR when increasing the channel count $N$, because the required ADC bandwidth $B = B_{opt}/(2N)$ is reduced and because lower-speed ADC typically offer a higher ENOB. In the following sections, we provide a quantitative estimate of the signal-to-noise-and-distortion ratio (SNDR) for different channel counts $N$. As some of the analyzed impairments through noise and distortion, such as signal-signal beat interference (SSBI) or ADC noise, depend on the shape of the input signal, we focus the following discussion on a broadband, noise-like random test signal that is defined in Eq. (11) below and that is expected to approximate typical technical signals much better than a simple sinusoidal. To clearly differentiate the noise and distortion impairments obtained for such broadband test signals from those obtained for pure sinusoidals, we use different symbols: SNDR refers to the case of broadband test signals as defined in Eq. (11) below, whereas the term SINAD is used in case of sinusoidal test signals as in Fig. 5 above. In the following, we find an approximate description for the bandwidth-dependent SINAD of conventional ADC, see Section VI A. Based on this, we then analyze the ADC noise and various other noise contributions and distortions in an OAWM system and investigated how the associated SNDR contributions scale with channel count $N$. We then estimate the SNDR levels that can be expected for OAWM systems with different channel counts $N$ and hence with different optical acquisition bandwidths $B_{opt} \approx N \times 2B$, Section VI B. We find that increasing the channel count $N$ can effectively reduce the impact of ADC noise and thereby improve the overall acquisition performance, Section VI C.

### A. Noise limitations of electronic ADC

Figure 6 shows the ENOB of commercially available high-speed ADC (marked with "x") [22-27] and oscilloscopes (marked with "o") [28-34] as a function of the usable electrical acquisition bandwidth $B$. Note that, for some ADC chips, the native analogue bandwidth exceeds the Nyquist frequency $f_s/2$, where $f_s$ is the sampling frequency of the device. To avoid aliasing, we assume an appropriate low-pass filter with bandwidth $B = f_s/2$ for these devices. Expectedly, the ENOB of high-speed ADC and oscilloscopes reduces with increasing acquisition bandwidths $B$, which can be attributed to thermal noise and to jitter of the underlying sampling clock [35-37]. Thermal noise has a flat power spectral density such that the associated SINAD scales inversely proportional with the analog acquisition bandwidth $B$ [35-37],

$$\text{SINAD}^{(\text{therm})} = \frac{C_1}{B}. \tag{9}$$

In the above equation, $C_1$ is a proportionality constant that depends on the temperature, the full-scale voltage of the ADC, and the overall noise properties of the ADC, often quantified by an effective noise resistance [36].

Besides thermal noise, the SINAD of ADC can be impaired by timing jitter. The overall timing jitter of an ADC can be separated into a contribution from intrinsic aperture jitter $\tau_a$, that is inherent to the ADC's design and arises from thermal noise in the internal clock buffers [38,39], and a contribution from clock jitter $\tau_{clk}$, that is caused by the phase noise of the clock source itself. As the phase noise typically follows a

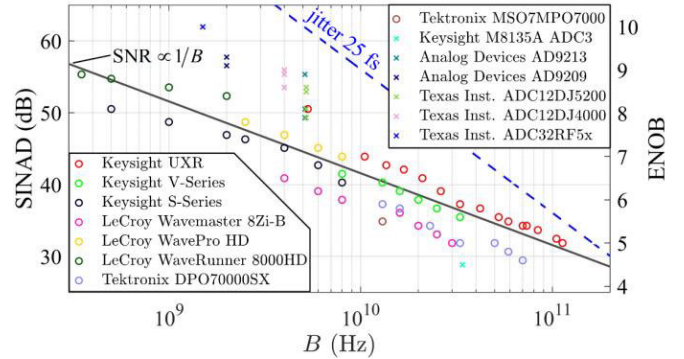

**Fig. 6.** Effective number of bits (ENOB) and associated signal-to-noise-and-distortion ratio (SINAD) as a function of the analog input bandwidth $B$ for different ADC (marked with "x") [22-27] and oscilloscopes (marked with "o") [28-34]. Even for the most advanced oscilloscopes, the SINAD decays approximately in proportion to $1/B$ (black solid line), see Eq. (9), indicating that the performance is rather limited by thermal noise than by jitter. The blue dotted line shows the jitter-limited SINAD according to Eq. (10) for the maximum frequency $f_{sig} = B$ and an RMS jitter $t_j$ of 25 fs, which is obtained for state-of-the-art real-time oscilloscopes in case of record lengths up to 10 μs [28]. The rms jitter of modern ADC chips of the order of 50 fs [23-27].

Wiener process, the observed rms clock jitter $\tau_{clk}$ depends strongly on the observation time. Note, however, that the ENOB of an ADC is typically measured using a clean input tone and a short observation time, such that a low-frequency drift of the clock does not impact the resulting SINAD [40]. As clock jitter and intrinsic aperture jitter are independent, their variances can be added to obtain the total rms-jitter $\tau_j = \sqrt{\tau_a{}^2 + \tau_{clk}{}^2}$. The jitter-limited SINAD of a sinusoidal test signal at frequency $f_{sig}$ is given by [41],

$$\text{SINAD}^{(\text{jitter})} = \frac{1}{\left(2\pi f_{sig}\tau_j\right)^2} \, . \tag{10}$$

To get a quantitative understanding of jitter-induced performance limitations of ADC, we have plotted the worst-case SINAD contribution for a sinusoidal signal with the highest possible frequency, $f_{sig} = B$, as a dashed blue line into Fig. 6, assuming an overall RMS jitter of 25 fs as obtained for state-of-the-art real-time oscilloscopes for record lengths up to 10 µs [28].

Our investigation shows that, for frequencies up to approximately 100 GHz, the practically achievable SINAD of high-end oscilloscopes is mainly limited by thermal noise and may hence be approximated by a rather simple $1/B$ relationship according to Eq. (9). As a quantitative model function for the subsequent analysis, we fit a $C_1/B$ curve to the bandwidth dependent SINAD values of the various high-speed ADC and oscilloscopes, where $C_1$ is the only free fitting parameter. The resulting fit ($C_1 \approx 150\,\text{THz}$) is indicated by a black trace in Fig. 6 and can approximate the real SINAD-bandwidth relationship reasonably well. It should also be noted that the use of low-jitter comb sources may even overcome the jitter limitations of all-electronic ADC, thereby paving the way towards high-SINAD acquisition at bandwidths well beyond 100 GHz [43].

### B. OAWM noise model

In the following, we discuss noise sources limiting the SNR, or more generally the SNDR for non-sliced OAWM systems featuring different numbers $N$ of parallel IQR, $N = 1,...,32$, and different optical acquisition bandwidths $B_{opt}$. For simplicity, we assume that the number of LO comb lines $M$ equals the number of IQ receiver channels $N$, $M = N$, and that the overlap region, see OR in Fig. 2, is small compared to the bandwidth $B$ of the individual IQR, such that the ADC bandwidth is approximately half the FSR of the LO comb, $B \approx f_{FSR}/2$. Depending on the optical acquisition bandwidth $B_{opt}$ and the channel count $N$, the required ADC bandwidth amounts to $B \approx B_{opt}/(2N)$. In our model, we consider acquisition noise from the ADC, which is assumed to scale according to the black trace in Fig. 6. We further account for shot noise of the photodetectors, thermal noise from electrical amplifiers, ASE noise from the optical amplifiers, jitter of the LO comb and the various ADC, as well as SSBI due to imperfect balancing of the photodetectors and errors introduced by calibration errors and the digital signal reconstruction. All modeled noise sources and distortions are visualized in the setup sketch in Fig. 7. We exclude the optical phase of the LO from this analysis because this limitation is independent of the optical acquisition bandwidth $B_{opt}$ and the channel count $N$. Note, however, that the optical phase noise of the LO comb will limit the maximum recording length in some applications. Phase-noise induced distortions may, e.g., be avoided by homodyne detection schemes as used in photonic-electronic ADC schemes [10-13], or by phase recovery algorithms in case of data signals.

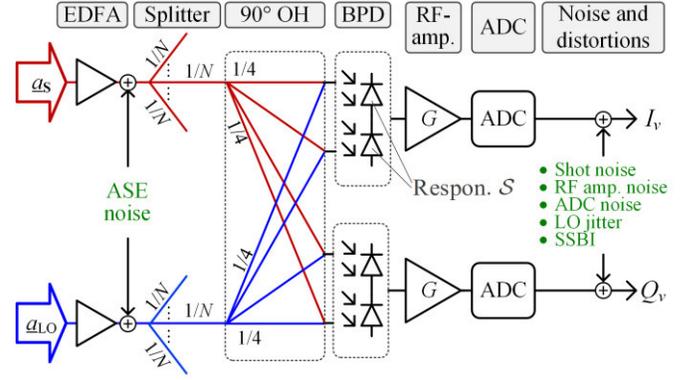

**Fig. 7.** Block diagram of a specific detection channel (index $\nu$) of the non-sliced OAWM system including different noise sources. The optical signal $\underline{a}_S$ and LO $\underline{a}_{LO}$ are first amplified by erbium-doped fiber amplifiers (EDFA). The EDFA add amplified spontaneous emission (ASE) noise. Subsequently, the signal and LO are split into $N$ copies. The copies of signal and LO are further split and combined by an 90° optical hybrid (OH) and fed to two pairs of balanced photodetectors (BPD) having responsivity $\mathcal{S}$. The generated RF output signals are amplified by RF amplifiers (gain $G$) and acquired by ADC. All further noise contributions, e.g., shot noise, RF amplifier noise, ADC noise, LO jitter noise, as well as distortions such as signal-signal beat interference (SSBI) are referred to the resulting output signals $I_\nu$ and $Q_\nu$.

### 1) Signal and system model

As an approximation of a real-world technical waveform, we assume a noise-like random test signal $\underline{a}_S(t) = \underline{A}_S(t)\exp\left(j2\pi f_{cntr}t\right)$, having a slowly varying complex-valued time-domain envelope $\underline{A}_S(t)$ modulated onto an optical carrier at frequency $f_{cntr}$. The amplitude $\underline{A}_S(t)$ is mean-free, and the associated real and imaginary parts are statistically independent and Gaussian distributed. We assume $\underline{A}_S(t)$ to have an average power of $P_S$ and constant double-sided power spectral density of $P_S/B_{opt}$ within the range $[-B_{opt}/2, B_{opt}/2]$, which corresponds to the acquisition range of our OAWM system. The real and the imaginary part of the test signal $\underline{A}_S(t)$ are essentially obtained by filtering two statistically independent spectrally white Gaussian noise processes with double-sided power spectral densities $P_S/(2B_{opt})$ by a low-pass with single-sided bandwidth $B_{opt}/2$. As a result, different spectral components of the corresponding amplitude spectrum $\underline{\tilde{A}}_S(f)$ are uncorrelated. Under these assumptions, the power spectral density $\tilde{S}_a(f)$ of the random optical signal $\underline{a}_S(t)$ centered at frequency $f_{cntr}$ can be written as

$$\tilde{S}_a(f) = \begin{cases} \dfrac{P_S}{B_{opt}} & \text{for } |f - f_{cntr}| \le \dfrac{B_{opt}}{2} \, , \\ 0 & \text{otherwise.} \end{cases} \tag{11}$$

We further assume for simplicity that all relevant system properties, e.g., the transfer functions $\underline{\tilde{H}}_{\nu\mu}^{(I,t)}(f)$ and $\underline{\tilde{H}}_{\nu\mu}^{(Q,t)}(f)$

as well as the common-mode rejection ratio (CMRR) are frequency-independent within the frequency band $-B \le f \le +B$ of the corresponding detection channel, and that the LO delays $\tau_\nu$ of the OAWM system are evenly distributed over the repetition period of the LO, which leads to the best possible conditioning of the transfer matrix $\underline{\tilde{\mathbf{H}}}(f)$ [8]. Under these assumptions, the frequency-independent transfer matrix $\underline{\tilde{\mathbf{H}}}(f)$ is unitary and consequently a multiplication of its inverse $\underline{\tilde{\mathbf{H}}}^{-1}(f)$ with a vector $\left[ \tilde{\mathbf{I}}^{\mathrm{T}}(f), \tilde{\mathbf{Q}}^{\mathrm{T}}(f) \right]^{\mathrm{T}}$ of noisy I- Q signals as in Eq. (5) leaves the SNR unchanged. The SNDR of the finally reconstructed signal is thus dictated by the SNDR of the individual in-phase $I_\nu$ and quadrature $Q_\nu$ signals as well as by additional impairments due to calibration and reconstruction errors, see crosstalk in Eq. (5) and discussion thereof. We may hence first derive an expression for the SNDR of one baseband signal, e.g., $I_1$, and later find the overall SNDR by adding the additional impairments introduced by the signal reconstruction.

To estimate the SNDR of a single baseband signal, we first derive expressions for the generated RF signal power $P_{\mathrm{RF,S}}$ in each BPD, see Fig. 7. To this end, we make again use of the simplifying assumption that the receiver transfer functions $\underline{\tilde{H}}_{\nu\mu}^{(\mathrm{I,t})}(f)$, see Eq. (1), are frequency-independent within the associated frequency band $-B \le f \le +B$, and express the modulus of the transfer functions using the physical system parameters in TABLE I, and Fig. 7.

$$\left| \underline{\tilde{H}}_{\nu\mu}^{(\mathrm{I,t})}(f) \right| =$$

$$\begin{cases} \underbrace{\sqrt{G R}}_{\substack{\text{RF gain} \\ \text{and ADC} \\ \text{impedance}}} \times \underbrace{2\mathcal{S}}_{\substack{\text{Responsiv.} \\ \text{of balanced} \\ \text{detector}}} \times \underbrace{\frac{1}{\sqrt{4}\sqrt{4}}}_{\substack{\text{Optical} \\ \text{hybrid} \\ \text{(splitting} \\ \text{signal | LO)}}} \times \underbrace{\frac{1}{\sqrt{N}\sqrt{N}}}_{\substack{\text{Power} \\ \text{splitter} \\ \text{signal | LO}}} \times \underbrace{\frac{\sqrt{P_{\mathrm{LO}}}}{\sqrt{N}}}_{\substack{\text{LO} \\ \text{amplitude} \\ \text{per comb} \\ \text{line}}} & \begin{array}{l} \text{for } |f| \le B \text{ and} \\ \forall \, \nu, \mu \in [1, N] \end{array} \\ \\ 0 & \text{otherwise.} \end{cases}$$

$$(12)$$

In this relation, $\mathcal{S}$ refers to the responsivity of a single photodetector within a balanced pair such that the responsivity of the BPD is given by $2\mathcal{S}$. A relation equivalent to Eq. (12) applies to the receiver transfer functions $\underline{\tilde{H}}_{\nu\mu}^{(\mathrm{Q,t})}(f)$ associated with the quadrature components. Note Eq. (12) neglects the excess insertion loss of the various passive components for simplicity, e.g., optical waveguides, power splitters, and RF connectors, see in Fig. 7 – these losses can be included in Eq. (12) by modifying the various transmission factors accordingly.

For the random input signal $\underline{a}_{\mathrm{S}}(t) = \underline{A}_{\mathrm{S}}(t) \exp\left( \mathrm{j} 2\pi f_{\mathrm{cntr}} t \right)$ with slowly-varying complex envelope $\underline{A}_{\mathrm{S}}(t)$ as described above, the band-limited frequency-down-shifted signal portions $\mathcal{F}^{-1}\left\{ \underline{\tilde{H}}_{\nu\mu}^{(\mathrm{I,t})}(f) \underline{\tilde{a}}_{\mathrm{S}}(f + f_\mu) \right\}$ and $\mathcal{F}^{-1}\left\{ \underline{\tilde{H}}_{\nu\mu}^{(\mathrm{I,t})*}(-f) \underline{\tilde{a}}_{\mathrm{S}}^*(-f + f_\mu) \right\}$, $\mu = 1, ..., M$ are statistically independent from one another and Gaussian distributed in the time domain. According to Eqs. (1) and (2) the superposition of these band-limited, frequency down-shifted signal portions finally leads to the baseband signals $I_\nu(t)$ and $Q_\nu(t)$ recorded at the various IQ receivers IQR$_\nu$, which are also Gaussian distributed and statistically independent. The double-sided power-spectral-density

$\tilde{S}_{\mathrm{Iv}}^{(\mathrm{DS})}(f)$ of the received baseband voltage signal $I_\nu(t)$ can be obtained by adding the various frequency-shifted power spectra $\tilde{S}_{\mathrm{a}}(f + f_\mu)$ and $\tilde{S}_{\mathrm{a}}(-f + f_\mu)$ associated with the frequency-shifted signal portions $\underline{\tilde{a}}_{\mathrm{S}}(f + f_\mu)$ and $\underline{\tilde{a}}_{\mathrm{S}}^*(-f + f_\mu)$,

$$\begin{aligned} \tilde{S}_{\mathrm{Iv}}^{(\mathrm{DS})}(f) &= \sum_{\mu=1}^{N} \frac{\left| \underline{\tilde{H}}_{\nu\mu}^{(\mathrm{I,t})}(f) \right|^2}{R} \left[ \tilde{S}_{\mathrm{a}}(f + f_\mu) + \tilde{S}_{\mathrm{a}}(-f + f_\mu) \right] \\ &= \sum_{\mu=1}^{N} 2 \frac{\left| \underline{\tilde{H}}_{\nu\mu}^{(\mathrm{I,t})}(f) \right|^2}{R} \tilde{S}_{\mathrm{a}}(f + f_\mu) \\ &= \begin{cases} \dfrac{G R \mathcal{S}^2 P_{\mathrm{LO}}}{2N^3} \sum_{\mu=1}^{N} \tilde{S}_{\mathrm{a}}(f + f_\mu) & \text{for } |f| \le B, \\ 0 & \text{otherwise.} \end{cases} \end{aligned}$$

$$(13)$$

Integrating $\tilde{S}_{\mathrm{Iv}}^{(\mathrm{DS})}(f)$ within the double-sided frequency interval $[-B, B]$ covered by the receiver leads to the average RF signal power $P_{\mathrm{RF,S}}$ of the received baseband voltage signal $I_\nu(t)$,

$$P_{\mathrm{RF,S}} = \int_{-B}^{B} \tilde{S}_{\mathrm{Iv}}^{(\mathrm{DS})}(f) \, \mathrm{d}f = \frac{B G R \mathcal{S}^2 P_{\mathrm{LO}} P_{\mathrm{S}}}{N^2 B_{\mathrm{opt}}} = \frac{G R \mathcal{S}^2 P_{\mathrm{LO}} P_{\mathrm{S}}}{2N^3}, \, (14)$$

where $B_{\mathrm{opt}} \approx N \times 2B$ is used for the last step. Note that the same result applies for the received quadrature signals $Q_\nu(t)$. We assume that the total optical input power $P_{\mathrm{S}}$ of the signal and the total LO power $P_{\mathrm{LO}}$ can be increased at will to compensate for the additional splitting loss such that the optical power incident on each photodetector $P_{\mathrm{PD}} = \left( P_{\mathrm{S}} + P_{\mathrm{LO}} \right) / \left( 4N \right)$ of each balanced pair becomes independent of the number of detection channels $N$. We can thus rewrite Eq. (14) by using the LO-to-signal-power ratio (LOSPR),

$$P_{\mathrm{RF,S}} = \frac{\mathrm{LOSPR}}{\left( 1 + \mathrm{LOSPR} \right)^2} \frac{8 G R \mathcal{S}^2 P_{\mathrm{PD}}^2}{N}, \quad \mathrm{LOSPR} = \frac{P_{\mathrm{LO}}}{P_{\mathrm{S}}}. \quad (15)$$

Note that the RF power of the electrical signals generated by the balanced photodetectors within the detector band $-B \le f \le +B$ decays in proportion to $1/N$, even if a constant photodetector input power $P_{\mathrm{PD}}$ is maintained, because an increasing fraction of the down-converted signal is outside the detector's bandwidth. Consequently, the gain $G$ of the RF amplifier following the photodetectors must be increased in proportion to $N$ such that the resulting RF power $P_{\mathrm{RF,S}}$ fed to the ADC becomes again independent of $N$. Based on our simplified model, the RF-amplifier gain should be increased from $G_{\mathrm{dB}} = 6 \, \mathrm{dB}$ for $N = 1$ to $G_{\mathrm{dB}} = 21 \, \mathrm{dB}$ for $N = 32$, to achieve a peak-to-peak voltage of 500 mV at the ADC's input.

To quantitatively estimate the SNDR, $\mathrm{SNDR}^{(\mathrm{total})}$, of the measured in-phase and quadrature signals $I_\nu$ and $Q_\nu$, we need to relate the signal power $P_{\mathrm{RF,S}}$ according to Eq. (15) to the sum of all noise contributions and all further signal distortions, as discussed in the following sections. We will use the term "SNDR contribution" to refer to the noise contribution of a certain noise source towards the overall noise level. Note that the overall SNDR is obtained by adding the reciprocal values of the associated "SNDR contributions" and taking the inverse of the resulting sum.

TABLE I. SYSTEM PARAMETERS

| Symbol | Quantity | Value |
|---|---|---|
| $R$ | Input impedance of RF amplifier and ADC | 50 Ω |
| $\mathrm{NF_{dB}}$ | Noise figure of RF amplifier $\mathrm{NF_{dB}} = 10\log_{10}(F)$ | 10 dB |
| $\mathcal{S}$ | Responsivity of single photodetector | 0.8 A/W |
| $P_{PD}$ | Total optical power per photodetector | 0 dBm |
| $\mathrm{CMRR_{dB}}$ | Common mode rejection ratio (CMRR) $\mathrm{CMRR_{dB}} = 20\log_{10}(\mathrm{CMRR})$ | -30 dB |
| $\mathrm{LOSPR_{dB}}$ | LO-to-signal power ratio $\mathrm{LOSPR_{dB}} = 10\log_{10}(P_{LO}/P_S)$ | 10 dB |
| $\tau_j^{(ADC)}$ | rms jitter of ADC | 25 fs |
| $\tau_j^{(LO)}$ | rms jitter of LO comb | 25 fs |
| $G_{dB}$ | Gain of RF amplifier: $G_{dB} = 10\log_{10}(G)$ | 6 dB to 21 dB |
| $\mathrm{OSNR_{sig}}$ | Optical signal-to-noise ratio of signal ($\Leftrightarrow$ input power $P_{in} \approx -13\,\mathrm{dBm}$) | 40 dB |

*2) Shot noise*

The shot-noise-related current variance $\overline{i_{shot}^2}$ in each of the two photodiodes of a single balanced detector is calculated from the average photocurrent $\overline{i_{ph}}$ and the elementary charge $e$,

$$\overline{i_{shot}^2} = 2e\overline{i_{ph}}\,B = \frac{e\,\overline{i_{ph}}B_{opt}}{N}, \qquad \overline{i_{ph}} = \mathcal{S} \times P_{PD}. \quad (16)$$

The shot-noise contributions of the two photodiodes in a balanced pair are statistically independent, and the associated powers hence add. The overall electrical shot-noise is further boosted by the RF amplifier with gain $G$, see Fig. 7, and then fed to the ADC with input impedance $R$. The output-referred shot-noise power $P_{N,shot}$ can hence be written as

$$P_{N,shot} = 2G\overline{i_{shot}^2}R = \frac{2G\mathcal{S}qP_{PD}B_{opt}R}{N}, \quad (17)$$

and the associated contribution of the shot noise to the overall SNDR is

$$\mathrm{SNDR}^{(shot)} = \frac{P_{RF,S}}{P_{N,shot}} = \frac{\mathrm{LOSPR}}{(1+\mathrm{LOSPR})^2}\frac{4\mathcal{S}P_{PD}}{qB_{opt}}. \quad (18)$$

We compare the SNDR associated with different noise sources and distortions by plotting the individual SNDR contributions as a function of the optical acquisition bandwidth $B_{opt}$ for channels counts $N = 1, 4, 16, 32$, see Fig. 8 (a), (b), (c), and (d). The shot-noise-related SNDR is shown as a dark gray dashed line and decays by 10 dB per decade due to the spectrally white nature of shot noise. Note that we assumed ADC bandwidths $B$ of up to 100 GHz, such that the maximum achievable optical acquisition bandwidth $B_{opt} \approx N \times 2B$ amounts to 200 GHz for $N = 1$, Fig. 8 (a). In Fig. 9 (a), (b), (c), and (d), we again use gray lines to show the same SNDR contributions as a function of the channel count $N = 1,...,32$ for fixed optical acquisition bandwidths $B_{opt} = 200$ GHz, 400 GHz, 800 GHz, 1 THz. As expected from Eq. (18), the resulting shot noise contribution to the SNDR is independent of the channel count $N$.

*3) RF-amplifier noise*

The signals generated by the photodetectors are fed to an RF amplifier, Fig. 7, which contributes thermal noise. We model the total output-referred thermal noise of the RF amplifier using its noise factor $F$,

$$P_{N,RF-amp} = GFkTB. \quad (19)$$

The SNDR contribution related to the RF-amplifier, $\mathrm{SNDR}^{(RF-amp)}$, is thus given by

$$\mathrm{SNDR}^{(RF-amp)} = \frac{P_{RF,S}}{P_{N,RF-amp}} = \frac{\mathrm{LOSPR}}{(1+\mathrm{LOSPR})^2}\frac{8R\mathcal{S}^2 P_{PD}^2}{FkTB_{opt}}. \quad (20)$$

In our quantitative model, we assume a noise figure of $\mathrm{NF_{dB}} = 10\log_{10}(F) = 10$ dB for the electrical amplifiers, see TABLE I. This leads to the SNDR contributions plotted as green dashed lines in Fig. 8 and Fig. 9. The $\mathrm{SNDR}^{(RF-amp)}$ as a function of the optical acquisition bandwidth $B_{opt}$ decays by 10 dB per decade over frequency due to the white power spectral density of thermal noise.

*4) Noise and distortions of the electronic ADC*

Next, we consider ADC noise, which we estimate from the bandwidth-dependent SINAD and ENOB values shown in Fig. 6. Note that the values indicated in Fig. 6 refer to sinusoidal test signals (PAPR = 2) such that the corresponding fit $\mathrm{SINAD} = C_1/B$, ($C_1 \approx 150$ GHz) must be re-scaled for test signals with different PAPR. To this end, we define a new proportionality constant $C_2 = (2/\mathrm{PAPR})C_1$, where PAPR refers to the peak-to-average power ratio of the test signal. It should be noted that for the broadband test signals defined in Eq. (11), the real and imaginary parts of the complex-valued time-domain amplitude follow independent Gaussian distributions with infinite tails, such that the peak amplitude and hence the PAPR are not well defined. To resolve this issue, we assume that all I/Q signals measured by the ADC are clipped at $\pm 4\sigma$, where $\sigma$ is the standard deviation of the respective Gaussian signal – corresponding to a ratio of clipped samples within the overall recording below $10^{-4}$. This limits the PAPR to 16 ($\approx$12 dB), which is 9 dB higher than that of a sinusoidal test-signal. Consequently, the $\mathrm{SNDR}^{(ADC)}$ related to the ADC is 9 dB lower than the associated SINAD plotted in Fig. 6,

$$\mathrm{SNDR}^{(ADC)} = \frac{C_2}{B} = \frac{C_2 N}{B_{opt}}, \qquad C_2 = \frac{2C_1}{\mathrm{PAPR}} \approx 19\,\mathrm{THz}. \quad (21)$$

We again indicate the SNDR contribution associated with the ADC in Fig. 8 and Fig. 9, respectively, using red dashed lines. For low channels counts $N$, the ADC noise is the dominant noise source, see Fig. 8(a) and (b).

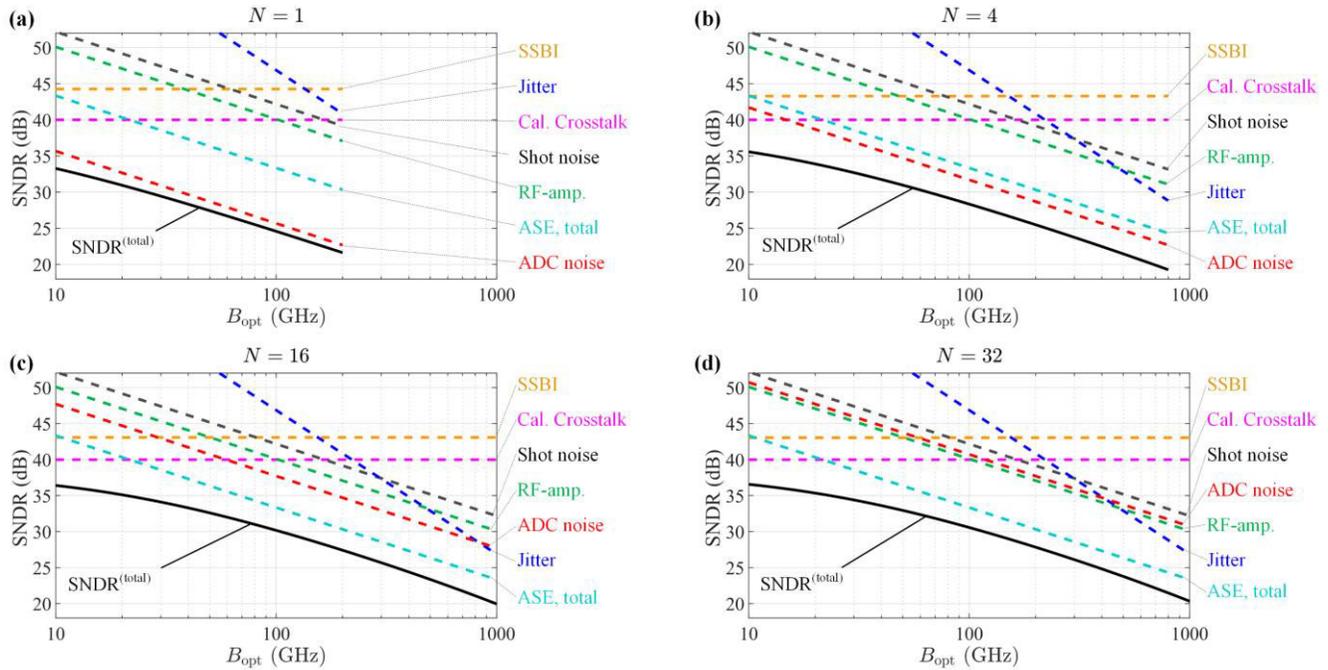

**Fig. 8.** Individual contributions of different sources of noise and distortion to the overall signal-to-noise-and-distortion ratio SNDR[(total)] (solid black lines) for a broadband test signal as a function of the optical acquisition bandwidth $B_{opt}$ for different channel counts $N = 1, 4, 16, 32$, which are represented by Subfigures (a), (b), (c), and (d). We consider shot noise (dashed black lines), RF-amplifier noise (dashed green lines), ADC noise (dashed red lines), ASE noise from signal and LO (dashed cyan lines), jitter of the ADC and the LO comb (dashed blue lines), signal-signal beat interference (SSBI dashed yellow lines), and calibration crosstalk (dashed magenta lines). We assume that the ADC bandwidth $B$ is limited to 100 GHz, such that the maximum achievable optical acquisition bandwidth $B_{opt} \approx N \times 2B$ amounts to 200 GHz for $N = 1$ and to 800 GHz, for $N = 2$, see Subfigures (a) and (b). The model assumes an OSNR for the signal of 40 dB ($P_{in} \approx -13\,\mathrm{dBm}$), an OSNR for the LO of 48 dB and further relies on the parameters specified in Table I and TABLE II. For low channel counts, the ADC noise dominates over other noise sources and represents the main performance limitation. Increasing the channel count $N$ allows to reduce the bandwidth of the individual ADC and thus improves the overall SNDR[(total)].

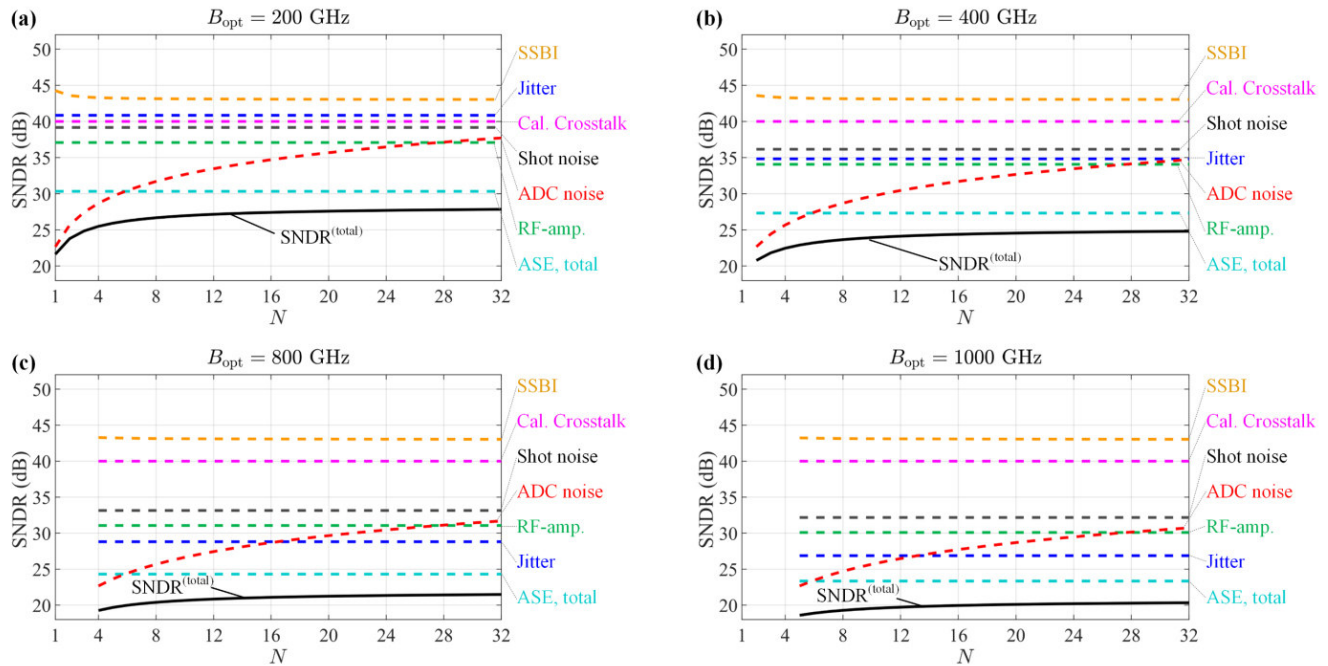

**Fig. 9.** Individual contributions of different sources of noise and distortion to the overall signal-to-noise-and-distortion ratio (SNDR[(total)], solid ,black lines) for a broadband test signal as a function of the channel count $N$ for different overall optical acquisition bandwidths $B_{opt}$ of 200 GHz, 400 GHz, 800 GHz, and 1 THz, which are represented by Subfigures (a), (b), (c), and (d). We consider shot noise (dashed, black lines), thermal RF-amplifier noise (dashed, green lines), ADC noise (dashed, red lines), ASE noise from signal and LO (dashed cyan lines), jitter of the ADC and the LO comb (dashed blue lines), signal-signal beat interference (SSBI, dashed yellow lines), and calibration crosstalk (dashed magenta lines). The model assumes an OSNR for the signal of 40 dB ($P_{in} \approx -13\,\mathrm{dBm}$), an OSNR for the LO of 48 dB, and further relies on the parameters specified in Table I and TABLE II. For low channel counts ($N < 6$), the noise of the electronic ADC dominates over other sources of noise and distortion sources, and the overall SNDR improves by parallelizing more lower-speed ADC with higher ENOB. However, the impact of larger channel counts $N$ becomes less significant for $N \geq 6$ because the ASE takes over as a dominant noise source. Further improvement would only be possible for input signals with even higher OSNR, e.g., $\mathrm{OSNR_{sig}} = 50\,\mathrm{dB}$ ($P_{in} = -3\,\mathrm{dBm}$) – in this case, scaling the channel count beyond $N = 6$ would be advantageous to sustain higher overall SNDR.

## 5) Jitter of LO comb and ADC clock

In addition, jitter of the LO comb and the ADC array may impair the acquired signals, in particular when investigating very broadband waveforms. In general, comb-based OAWM schemes are subject to jitter-induced impairments originating from both the electronic ADC clock and the LO comb [10]. Assuming both jitter processes to be statistically independent, the noise-to-signal ratios (NSR) induced by ADC jitter $\tau_{\mathrm{j}}^{(\mathrm{ADC})}$ and LO-comb jitter $\tau_{\mathrm{j}}^{(\mathrm{LO})}$ can be added to calculate the combined NSR. In our analysis we follow the procedure described for a photonic-electronic ADC in [10] and adapt it to the case of our non-sliced OAWM system, see Appendix A for details. Using this approach, the jitter-related contribution to the overall SNDR of a broadband test signal can be written as

$$
\begin{aligned}
\mathrm{SNDR}^{(\mathrm{jitter})} &= \frac{12 N^2}{\left(2\pi B_{\mathrm{opt}}\right)^2 \left[\tau_{\mathrm{j}}^{(\mathrm{ADC})\,2} + \tau_{\mathrm{j}}^{(\mathrm{LO})\,2}\left(N^2-1\right)\right]} \\
&= \frac{3}{\left[2\pi\left(B_{\mathrm{opt}}/2\right)\tau_{\mathrm{j}}\right]^2} \quad \text{for } \tau_{\mathrm{j}} = \tau_{\mathrm{j}}^{(\mathrm{ADC})} = \tau_{\mathrm{j}}^{\mathrm{LO}}.
\end{aligned}
\tag{22}
$$

For simplicity, we further assume that the LO comb and the RF comb have the same rms-jitter $\tau_{\mathrm{j}} = \tau_{\mathrm{j}}^{(\mathrm{ADC})} = \tau_{\mathrm{j}}^{(\mathrm{LO})}$, which may, be achieved by either generating the LO comb through modulation of a continuous-wave (CW) laser tone or by using RF-injection-locked of the Kerr combs [44]. Note that this relation refers to the averaged jitter-related SNDR of a spectrally white broadband optical test signal and thus differs from Eq. (10), which gives the jitter-related SINAD of sinusoidal electrical test tone that is acquired by a certain ADC.

The resulting $\mathrm{SNDR}^{(\mathrm{jitter})}$ is plotted as a function of the optical acquisition bandwidth $B_{\mathrm{opt}}$ for $\tau_{\mathrm{j}} = 25\,\mathrm{fs}$ in Fig. 8. As expected from Eq. (22), the SNDR decays by 20 dB per decade over frequency, but only starts to play a role for bandwidths above 500 GHz. These limitations could be overcome by ultra-low jitter levels as offered by so-called self-injection locked Kerr frequency combs, which have the potential to outperform high-quality electronic oscillators in the future [45]. When plotted as a function of the channel count $N$ and for a fixed optical acquisition bandwidth $B_{\mathrm{opt}}$, see Fig. 9, the $\mathrm{SNDR}^{(\mathrm{jitter})}$ remains independent of $N$ for the assumption $\tau_{\mathrm{j}} = \tau_{\mathrm{j}}^{(\mathrm{ADC})} = \tau_{\mathrm{j}}^{(\mathrm{LO})}$ contained in Eq. (22). Note that, in case the LO-comb jitter is lower that the ADC jitter, $\tau_{\mathrm{LO}} < \tau_{\mathrm{ADC}}$, increasing the channel count $N$ might also help to mitigate the impact of electronic jitter [10].

## 6) ASE noise

Next, we consider the noise contributions of the optical amplifiers. The signal amplifier, typically implemented as an EDFA, see Fig. 7, allows to operate the OAWM system over a wide range of signal input powers by adjusting the power $P_{\mathrm{PD}}$ that is received by the various photodetectors. However, this optical amplifier inevitably adds amplified spontaneous emission (ASE) noise, which limits the optical signal-to-noise ratio (OSNR) of the amplified signal [46]. The OSNR is typically measured in a standard reference bandwidth of $B_{\mathrm{ref}} = 12.5\,\mathrm{GHz}$ (0.1 nm at a center wavelength of 1550 nm) using an optical signal analyzer (OSA) that collects the ASE noise in both polarizations. For a quantitative estimate of the ASE-related impairments, we characterize the OSNR generated by the erbium-doped fiber amplifier (EDFA) used in our experiments (Connect, MFAS-ER-C-M-20-PA). To this end, we vary the optical input power $P_{\mathrm{in}}$ and the pump current of the device and record the resulting OSNR and the associated output power $P_{\mathrm{out}}$. The results are shown in Fig. 10 (a), where the output power $P_{\mathrm{out}}$ is color-coded. We find that, for sufficiently high input power levels above $-30\,\mathrm{dBm}$, the OSNR at the output is essentially independent of the output power level. Thus, compensating the splitting loss of a multi-channel OAWM system by further amplifying the input does not lead to an additional OSNR penalty. To convert the OSNR to an $\mathrm{SNDR}_{\mathrm{sig}}^{(\mathrm{ASE})}$ contribution for a given optical acquisition bandwidth $B_{\mathrm{opt}}$, we need to account for the fact that our OAWM scheme relies on a single-polarization receiver, whereas the OSNR considers ASE noise in both polarizations. This leads to the relation

$$
\mathrm{SNDR}_{\mathrm{sig}}^{(\mathrm{ASE})} = 2 \times \mathrm{OSNR} \times \frac{B_{\mathrm{ref}}}{B_{\mathrm{opt}}}.
\tag{23}
$$

Figure 10 (b) shows the corresponding color-coded contour plot, indicating the ASE-related SNDR contribution $\mathrm{SNDR}_{\mathrm{sig}}^{(\mathrm{ASE})}$ as a function of the OSNR and the optical acquisition bandwidth $B_{\mathrm{opt}}$ ranging from 100 GHz to 1 THz. Expectedly, the impact of ASE noise on the resulting SNDR becomes more severe as the optical acquisition bandwidth increases.

It should also be noted that in many cases, the LO comb also requires optical amplification, e.g., if the individual comb lines are too low in power to serve as LO tones. This leads to an ASE-related SNDR contribution $\mathrm{SNDR}_{\mathrm{LO}}^{(\mathrm{ASE})}$ for the LO, similar to the one shown for the signal in Fig. 10(b). To obtain the overall ASE-related SNDR contribution $\mathrm{SNDR}_{\mathrm{tot}}^{(\mathrm{ASE})}$ that accounts for the amplifier noise both in the signal and in the LO path, the two contributions need to be merged by adding the statistically independent ASE noise powers,

$$\text{SNDR}_{\text{tot}}^{(\text{ASE})} = \left[ \frac{1}{\text{SNDR}_{\text{signal}}^{(\text{ASE})}} + \frac{1}{\text{SNDR}_{\text{LO}}^{(\text{ASE})}} \right]^{-1}. \quad (24)$$

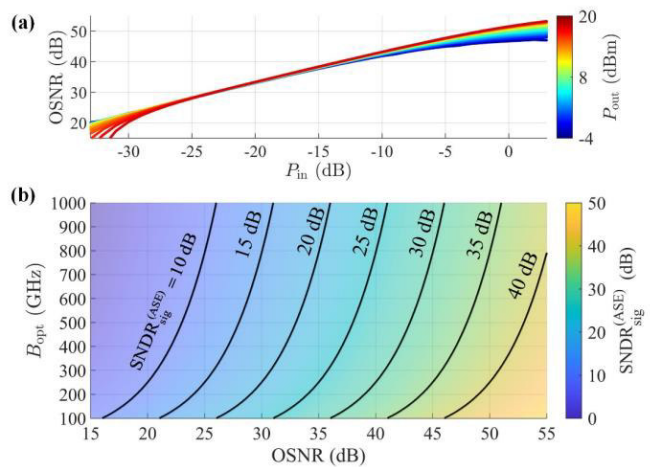

**Fig. 10**. SNDR contributions of ASE noise resulting from optical amplifiers. **(a)** Measured OSNR as a function of the optical input power $P_{\mathrm{in}}$ for various (color coded) output power levels $P_{\mathrm{out}}$ for the erbium-doped fiber amplifier (EDFA) labeled "AMP" in Fig. 1. For sufficiently high input power levels above -30 dBm, the OSNR at the output is essentially independent of the output power levels. **(b)** Color-coded contour plot according to Eq. (23), indicating the ASE-related SNDR contribution $\mathrm{SNDR}_{\mathrm{sig}}^{(\mathrm{ASE})}$ as a function of the OSNR and the optical acquisition bandwidth $B_{\mathrm{opt}}$, which ranges from 100 GHz to 1 THz. Expectedly, the impact of ASE noise on the resulting SNDR becomes more severe as the optical acquisition bandwidth increases.

Note that the ASE-related SNDR contribution $\text{SNDR}_{\text{LO}}^{(\text{ASE})}$ of the LO comb can be reduced by spectral filtering of the individual comb lines to suppress the ASE noise in-between. For our experiments, the four-tone LO features an OSNR of 48.5 dB, which after spectral filtering leads to an SNDR limit of approximately $\text{SNDR}_{\text{LO}}^{(\text{ASE})} = 42.4\,\text{dB}$. This is slightly higher than the 40.2 dB that would be expected from Eq. (23) for an optical acquisition bandwidth $B_{\text{opt}} = 170\,\text{GHz}$ without any additional noise suppression.

In Fig. 8 we plot the overall ASE-related SNDR contribution $\text{SNDR}_{\text{tot}}^{(\text{ASE})}$, dashed cyan lines, as a function the optical acquisition bandwidth $B_{\text{opt}}$, assuming an OSNR for the signal of 40 dB, which corresponds to an input power of $P_{\text{in}} = -13\,\text{dBm}$, and an OSNR for the LO comb of 48 dB, as measured for our demonstration experiment. Similarly, Fig. 9 shows the $\text{SNDR}_{\text{tot}}^{(\text{ASE})}$ as a function of the channel count $N$ for the same parameters. For large optical acquisition bandwidths, the ASE-noise-related SNDR contribution $\text{SNDR}_{\text{tot}}^{(\text{ASE})}$ dominates over other noise sources and dictates the achievable signal quality. Clearly, increasing the input power of the signal by 10 dB will decrease the ASE-related noise contribution of the signal by 10 dB and increase the resulting SNDR limit $\text{SNDR}_{\text{tot}}^{(\text{ASE})}$ accordingly, see Eq. (24).

### 7) Signal-signal beat interference (SSBI)

Besides the electrical and optical noise sources discussed above, the received in-phase and quadrature signals $I_\nu$ and $Q_\nu$ are subject to SSBI, which occurs as a consequence of imperfectly balanced photodetectors (BPD). As a metric for balancing, we use the effective common-mode rejection ratio (CMRR) of our detection system, which is based on the differences of the effective responsivities $\mathcal{S}_1$ and $\mathcal{S}_2$ of the two photodetectors within a given BPD. These effective responsivities do not only account for differences of the responsivities of the photodetectors themselves, but also for uneven splitting ratios of the respective optical hybrids as well as for differences of the optical and electrical transmission paths associated with the respective photodetector. For simplicity, we assume a frequency-independent CMRR that is given by [47],

$$\text{CMRR} = \frac{|\mathcal{S}_1 - \mathcal{S}_2|}{|\mathcal{S}_1 + \mathcal{S}_2|} = \frac{\Delta\mathcal{S}}{2\mathcal{S}}, \quad \text{CMRR}_{\text{dB}} = 20\log_{10}\left(\text{CMRR}\right), (25)$$

where $\Delta\mathcal{S} = |\mathcal{S}_1 - \mathcal{S}_2|$ denotes the difference of the effective responsivities and $\mathcal{S} = (\mathcal{S}_1 + \mathcal{S}_2)/2$ the associated average. The actual impairment by SSBI is naturally signal-dependent, and we again assume a random Gaussian-distributed test-signal with constant power spectral density within the optical acquisition bandwidth $B_{\text{opt}}$ of our OAWM system, see Eq. (11). In this case, the single-sided RF power spectrum $\tilde{S}_{\text{SSBI}}(f)$ associated with the SSBI has a triangular shape [48],

$$\tilde{S}_{\text{SSBI}}(f) = 2GR\Delta\mathcal{S}^2 \left( \frac{P_{\text{PD}}}{(1+\text{LOSPR})B_{\text{opt}}} \right)^2 \left( B_{\text{opt}} - f \right). (26)$$

In this relation, LOSPR refers to the LO-to-signal power ratio as defined for Eq. (15) above, $P_{\text{PD}}$ is again the total power incident on each of the two photodetectors that form the BPD, and $G$ and $R$ are the gain of the RF amplifier and the input impedance of the subsequent acquistion system, respectively. Note that, in case of limited CMRR, the output signal might also be impaired by LO-LO beat interference. The associated signal components, however, will only appear at zero frequency, leading to a DC offset of the photocurrent, and at the FSR frequency $f_{\text{FSR}}$ of the LO comb, which is not any more captured by the subsequent acquisition system, having a bandwidth $B$ slightly bigger than $f_{\text{FSR}}/2$. The overall SSBI power $P_{\text{SSBI}}$ within the receiver bandwidth $B \approx B_{\text{opt}}/(2N)$ is then obtained by integrating the single-sided power spectrum $\tilde{S}_{\text{SSBI}}(f)$ over the bandwidth of the respective receiver,

$$P_{\text{SSBI}} = \int_0^B \tilde{S}_{\text{SSBI}}(f)\,\text{d}f$$
$$= 2GR\Delta\mathcal{S}^2 \left( \frac{P_{\text{PD}}}{(1+\text{LOSPR})B_{\text{opt}}} \right)^2 \left( B_{\text{opt}}B - \frac{1}{2}B^2 \right) \quad (27)$$
$$= \frac{GR\Delta\mathcal{S}^2 P_{\text{PD}}{}^2 (4N-1)}{\left(1+\text{LOSPR}\right)^2 4N^2}.$$

Combining this relation with Eq. (15), we can write the SSBI-related SNDR contribution,

$$\text{SNDR}^{(\text{SSBI})} = \frac{P_{\text{RF,S}}}{P_{\text{SSBI}}} = \text{LOSPR}\left( \frac{1}{\text{CMRR}} \right)^2 \frac{8N}{(4N-1)}. \quad (28)$$

While for our model and the assumed parameters, $\text{CMRR}_{\text{dB}} = 30\,\text{dB}$ and $\text{LOSPR}_{\text{dB}} = 10\log_{10}\left(P_{\text{LO}}/P_{\text{S}}\right) = 10\,\text{dB}$, TABLE I, the SNDR contribution associated with SSBI is negligible, see dashed yellow lines in Fig. 8 and Fig. 9 above, the distortion becomes significant for a lower CMRR. Note that the observed SNDR reduces only slightly when increasing the number of parallel channels, Fig. 9. This can be explained by the SSBI-related power spectrum $\tilde{S}_{\text{SSBI}}(f)$ of the photocurrent, which, for a spectrally white optical signal, assumes a triangular shape, peaking at zero frequency (DC) and having a single-sided width of $B_{\text{opt}}$. For a single channel, $N = 1$, the associated noise power corresponds to the overall power contained in this triangular spectrum up to the detection bandwidth $B \approx B_{\text{opt}}/(2N) = B_{\text{opt}}/2$, whereas for $N \to \infty$, only the strong contributions close to the peak of the triangle at zero frequency play a role. As a result, the associated SNDR reduces by a factor of $3/4$ ($-1.25\,\text{dB}$) when increasing the channel count from $N = 1$ to $N \to \infty$, which can be seen by a slight initial decrease of the yellow curve in Fig. 9 (a).

### 8) Calibration and reconstruction errors

In addition, calibration crosstalk $\tilde{\underline{\mathbf{A}}}_{\text{X}}^{(\text{est})}(f)$ is introduced when reconstructing the signal-vector $\tilde{\underline{\mathbf{A}}}_{\text{S}}^{(\text{est})}(f)$ from the various measured baseband spectra $\tilde{I}_v(f)$ and $\tilde{Q}_v(f)$, because the reconstruction matrix is not known with perfect accuracy, see Eq. (5). The uncertainties in the reconstruction matrix arise on the one hand from uncertainties in the time-invariant transfer functions $\underline{\tilde{H}}_{v\mu}^{(1)}(f)$ and $\underline{\tilde{H}}_{v\mu}^{(Q)}(f)$, see Fig. 3, and, on the other hand, from estimation errors of the time variant factors $\underline{H}_{F,v}^{(1)} = \exp(j\varphi_{F,v}^{(1)})$, see Eq. (6). As these effects are rather complicated to describe analytically, we perform simulations to determine the SNDR contributions for different channel counts $N$. For these simulations, we assume relative uncertainties $10\log_{10}\left(\left|\Delta\underline{\tilde{H}}_{v\mu}^{(1)}(f)\right|^2 / \left|\underline{\tilde{H}}_{v\mu}^{(1)}(f)\right|^2\right) = -40$ dB and $10\log_{10}\left(\left|\Delta\underline{\tilde{H}}_{v\mu}^{(Q)}(f)\right|^2 / \left|\underline{\tilde{H}}_{v\mu}^{(Q)}(f)\right|^2\right) = -40$ dB for the frequency-dependent transfer functions $\underline{\tilde{H}}_{v\mu}^{(1)}(f)$ and $\underline{\tilde{H}}_{v\mu}^{(Q)}(f)$, which were estimated from the crosstalk levels observed in our experiments, see Fig. 5, and which are assumed to be independent of the channel count $N$. It should be noted that the quality of the calibration can be increased by averaging several calibration measurements at the expense of an increased one-time calibration effort. However, this will not permit to decrease the relative uncertainties of the frequency-dependent transfer functions $\underline{\tilde{H}}_{v\mu}^{(1)}(f)$ and $\underline{\tilde{H}}_{v\mu}^{(Q)}(f)$ at will due to remaining systematic errors, which will eventually limit the quality of the calibration. Our simulations are therefore based on an uncertainty level of -40 dB that was achieved in our proof-of-concept experiments.

For the simulation of the calibration crosstalk, we model the uncertainties $\Delta\underline{\tilde{H}}_{v\mu}^{(1)}(f)$ and $\Delta\underline{\tilde{H}}_{v\mu}^{(Q)}(f)$ as white Gaussian noise that distorts the various transfer-matrix elements, and we use again the broadband Gaussian test signal, Eq. (11), having an optical bandwidth of $B_{opt} = N \times f_{FSR}$, to evaluate the system performance. We run the simulation for a fixed ADC bandwidth of $B = 21$ GHz that is slightly larger than half the FSR, $f_{FSR}/2 = 20$ GHz, of the simulated LO comb, such that the time variant factors $\underline{H}_{F,v}^{(1)}$ and $\underline{H}_{LO,\mu}^{(1)}$ can be estimated. For our simulations, we rely again on the general system parameters shown in TABLE I, as used for the analytical estimations discussed above, along with the additional parameters listed TABLE II. While sweeping the channel count $N$, each simulation is repeated for different realizations of the noise processes, see Fig. 11 for the results. We extract and separately analyze the SNDR contributions related to the uncertain frequency-dependent transfer functions $\underline{\tilde{H}}_{v\mu}^{(1)}(f)$, $\underline{\tilde{H}}_{v\mu}^{(Q)}(f)$ (red trace) and the SNDR contribution related to estimation errors of the slowly time-variant factors $\underline{H}_{F,v}^{(1)}$ and $\underline{H}_{LO,\mu}^{(1)}$ (blue trace). We find that the SNDR contribution of the uncertain frequency dependent transfer functions $\underline{\tilde{H}}_{v\mu}^{(1)}(f)$, $\underline{\tilde{H}}_{v\mu}^{(Q)}(f)$ is independent of $N$ and amounts to approximately 40 dB, which corresponds approximately to the inverse of the underlying relative uncertainty $-10\log_{10}\left(\left|\Delta\underline{\tilde{H}}_{v\mu}^{(1)}(f)\right|^2 / \left|\underline{\tilde{H}}_{v\mu}^{(1)}(f)\right|^2\right) = 40$ dB of the transfer functions. In contrast to that, the SNDR contribution of the uncertain slowly time-variant complex-valued factors $\underline{H}_{F,v}^{(1)}$ and $\underline{H}_{LO,\mu}^{(1)}$ is significantly lower and does not seem to represent a relevant limitation of the overall signal quality. Note that the calibration crosstalk related to the frequency-dependent transfer functions does not vanish for $N=1$, see red trace in Fig. 11. This can be understood by considering that the IQ calibration crosstalk, i.e., the crosstalk between positive and negative spectral components, remains, see Fig. 5 for a measurement of the different crosstalk contributions.

TABLE II
SIMULATION PARAMETERS

| Symbol | Quantity | Value |
|---|---|---|
| $10\log_{10}\left[\dfrac{\left|\Delta\underline{\tilde{H}}_{v\mu}^{(1)}(f)\right|^2}{\left|\underline{\tilde{H}}_{v\mu}^{(1)}(f)\right|^2}\right]$ | Relative calibration errors | -40 dB |
| $f_{FSR}$ | Free spectral range of LO comb | 40 GHz |
| $B$ | BPD, RF-amplifier, and ADC bandwidth | 21 GHz |
| $NF_{dB}^{(EDFA)}$ | EDFA noise figure | 5 dB |
| $P_S$ | Signal input power | -12 dBm |
| $f_{sampling}^{(sim)}$ | Sampling rate of simulation | 1 THz |
| $N_{samples}$ | Number of simulated samples | $5\times10^6$ |

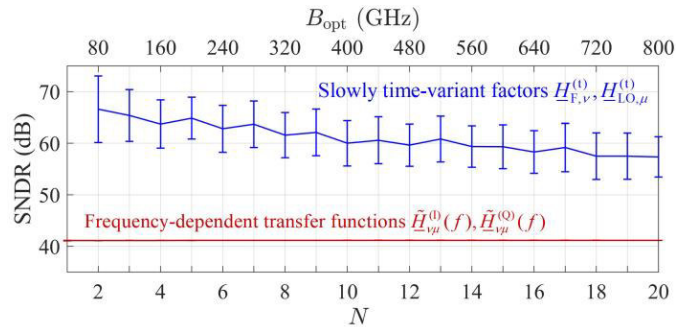

**Fig. 11.** Simulated SNDR contributions related to the uncertain frequency-dependent transfer functions $\underline{\tilde{H}}_{v\mu}^{(1)}(f)$, $\underline{\tilde{H}}_{v\mu}^{(Q)}(f)$ (red trace) and to the estimation errors of the slowly time-variant factors $\underline{H}_{F,v}^{(1)}$ and $\underline{H}_{LO,\mu}^{(1)}$ (blue trace) for increasing channel counts $N$. For the frequency-dependent transfer functions $\underline{\tilde{H}}_{v\mu}^{(1)}(f)$ and $\underline{\tilde{H}}_{v\mu}^{(Q)}(f)$, we assumed relative uncertainties $10\log_{10}\left(\left|\Delta\underline{\tilde{H}}_{v\mu}^{(1)}(f)\right|^2 / \left|\underline{\tilde{H}}_{v\mu}^{(1)}(f)\right|^2\right) = 10\log_{10}\left(\left|\Delta\underline{\tilde{H}}_{v\mu}^{(Q)}(f)\right|^2 / \left|\underline{\tilde{H}}_{v\mu}^{(Q)}(f)\right|^2\right) = -40$ dB that are independent of the channel count $N$ and that correspond to the uncertainty levels observed in our experiments. The underlying system and simulation parameters are summarized in TABLE I and TABLE II. The calibration crosstalk related to estimation errors of the slowly time-variant factors $\underline{H}_{F,v}^{(1)}$ and $\underline{H}_{LO,\mu}^{(1)}$ is much smaller than the contributions of the uncertain frequency-dependent transfer functions $\underline{\tilde{H}}_{v\mu}^{(1)}(f)$, $\underline{\tilde{H}}_{v\mu}^{(Q)}(f)$, indicated by a higher associated SNDR. Note that the calibration crosstalk related to the frequency-dependent transfer functions (red trace) does not vanish for $N=1$, because also this case is subject to crosstalk between positive and negative spectral components due to remaining IQ imbalance.

For comparison to the other noise and distortion sources, we include the overall calibration-related SNDR contribution $\mathrm{SNDR}^{(cal)} \approx 40$ dB into Fig. 8 and Fig. 9 above, see magenta traces. Under the assumptions explained above, this contribution neither changes with the overall optical acquisition bandwidth $B_{opt}$ nor with the channel count $N$ and does not represent a relevant limitation of the overall SNDR.

### C. Discussion of fundamental system limitations

We finally compare the different SNDR contributions analyzed in the previous section and calculate the resulting total $\mathrm{SNDR}^{(total)}$ that dictates the performance of the overall OAWM system. In Fig. 8 (a), (b), (c) and (d) above, we plotted all SNDR contributions as well as the resulting total $\mathrm{SNDR}^{(total)}$ (solid black lines) as a function of the optical acquisition bandwidth $B_{opt} \approx N \times 2B$ in the range of 10 GHz to 1 THz for different channel counts $N = 1, 4, 16, 32$, assuming that the electronic bandwidths $B$ of the underlying ADC can scale up to 100 GHz. Figure 9 (a), (b), (c), and (d) show the SNDR associated with the individual noise sources as well as the resulting total $SNDR^{(total)}$ as a function of the channel count $N$ for fixed optical acquisition bandwidths of 200 GHz, 400 GHz, 800 GHz, and 1 THz, respectively. We observe that for the assumed optical signal input power $P_{in} = -13$ dBm and the associated OSNR of 40 dB, the noise and distortions of the underlying electronic ADC dominate the overall impairments over a wide frequency range. Increasing the channel count $N$ allows to reduce the bandwidth of these electronic ADC and thus improves the overall $SNDR^{(total)}$ until other noise sources such as the ASE noise of the optical amplifiers becomes dominant. For the assumptions made in our model, ASE noise and ADC noise are equally strong at a channel count of approximately $N = 6$, see Fig. 9 (a), (b), (c) and (d). Within the parameter ranges considered here, jitter does not yet become a limitation. Note, however, that the underlying rms-jitter of 25 fs for the ADC was specified for an acquisition time of 10 µs [28]. Longer acquisition times might lead to a larger rms-jitter or require additional synchronization between the signal source and the OAWM system to avoid degradation of the overall $SNDR^{(total)}$, especially at large optical acquisition bandwidths $B_{opt}$. The RF-amplifier noise and the shot noise will only contribute significantly if both the ADC noise and the ASE noise limitations are overcome, e.g., in case of high optical input powers in combination with large channel counts and higher-ENOB ADC. The effect of SSBI and calibration crosstalk are negligible for the parameters assumed in our model.

We conclude that scaling up the channel count $N$ can not only rise the achievable optical acquisition bandwidth $B_{opt}$ with linearly increasing hardware effort, but also effectively reduce ADC noise and thus improve the overall SNDR. This emphasizes the potential of the presented approach for acquisition of ultra-broadband optical waveforms with high fidelity.

## VII. Summary

We have demonstrated non-sliced optical arbitrary waveform measurement (OAWM) with an acquisition bandwidth of 170 GHz, using a silicon PIC with co-integrated photodetectors. To the best of our knowledge, these experiments represent the first OAWM demonstration with an optical front-end having co-integrated photodetectors. In contrast to earlier OAWM schemes based on spectral slicing, our approach neither requires sensitive high-quality optical filters nor active controls. We show that the architecture reduces the bandwidth requirement for the individual ADC and improves the overall signal quality, because lower-speed ADC typically offer a higher effective number of bits (ENOB). Our system relies on a highly accurate calibration that uses a precisely defined femtosecond laser pulse as a known reference waveform. We obtain a signal-to-noise-and-distortion ratio of

30 dB when measuring single tone-test signals. This corresponds to an effective number of bits (ENOB) of 4.7 bit, where the underlying electronic analog-to-digital converters (ADC) turn out to be the main limitation. In a proof-of-concept experiment, we demonstrate the reception of 64QAM data signals at 100 GBd and obtain a constellation signal-to-noise ratio (CSNR) that is even slightly better than that achieved for a single intradyne IQ receiver based on discrete high-end 100 GHz photodiodes. We finally performed a theoretical scalability analysis, showing that the optical acquisition bandwidth and the signal quality can be further improved by increasing the channel count. The underlying model considers a broad range of additional other noise sources such as ASE noise in the signal and LO path, thermal RF-amplifier noise or jitter of the LO comb and the underlying electronic ADC, all of which can impact the performance of the OAWM system, especially for large optical acquisition bandwidths. The work paves the way towards further scaling of comb-based OAWM systems and is a key step towards out-of-lab use of highly compact OAWM systems that rely on chip-scale filter-less detector circuits.

## Appendix A

In this appendix, we derive Eq. (22) for the combined LO and ADC jitter observed for the broadband test signal defined in Eq. (11). As there are two independent contributions, the ADC jitter $\tau_j^{(ADC)}$ and LO-comb jitter $\tau_j^{(LO)}$, we first derive separate expressions for the individual noise-to-signal ratios (NSR), and add them in a second step according to Eq. 25 in [10] to obtain the overall NSR.

We first consider the electrical ADC with bandwidth $B$. As explained in Section VI A, the ADC jitter $\tau_j^{(ADC)}$ comprises both clock jitter and aperture jitter. We further assume that that the spectral components of the broadband test signal are uncorrelated. As a result, the NSR related to ADC jitter can be derived by taking the inverse jitter-related signal-to-noise-and-distortion ratio (SINAD) from Eq. (10) and by averaging it over the spectral range covered by the acquisition bandwidth $B$. This leads to

$$NSR_{ADC}^{(jitter)} = \frac{1}{B} \int_{f=0}^{B} \left(2\pi f \tau_j^{(ADC)}\right)^2 df$$

$$= \frac{\left(2\pi B \tau_j^{(ADC)}\right)^2}{3} = \frac{\left(2\pi B_{opt} \tau_j^{(ADC)}\right)^2}{12N^2}, \quad (29)$$

where the overall optical acquisition bandwidth $B_{opt} \approx N \times 2B$ was introduced in the last step for better comparison with other relations in Section VI.

For the LO comb, we consider the timing-jitter of the associated pulse train. Timing jitter contributes to the phase noise of the various comb tones, and the power of this contribution increases quadratically with frequency-difference of the respective tone and the carrier. In the subsequent analysis, we neglect optical phase noise of the carrier, see related discussion in Section VI.B, and we define a optical carrier frequency $f_{\mathrm{cntr}}$ to be in the center of the optical acquisition bandwidth $B_{\mathrm{opt}}$, see, e.g., Fig. 4. The effective jitter-related RF phase-noise power then also increases quadratically with the distance of the comb tone from the reference frequency $f_{\mathrm{cntr}}$ [49], and the NSR of the $\mu$-th LO comb line at frequency $f_\mu$ can be written as

$$\mathrm{NSR}^{(\mathrm{jitter})}_{\mathrm{LO},\mu} = \left[ 2\pi \left( f_\mu - f_{\mathrm{cntr}} \right) \tau^{(\mathrm{LO})}_{\mathrm{j}} \right]^2 . \quad (30)$$

Because we assume a broadband random test signal, see Eq.(11), all down-converted signal portions are statistically independent, even if the RF-phase of the individual tones are originally still correlated. Using this consideration, the overall NSR related to LO-comb jitter can be obtained by averaging the noise contributions associated with the $N$ individual comb tones,

$$\mathrm{NSR}^{(\mathrm{jitter})}_{\mathrm{LO}} = \frac{1}{N} \sum_{\mu=1}^{N} \mathrm{NSR}^{(\mathrm{jitter})}_{\mathrm{LO},\mu} = \frac{\left( 2\pi B_{\mathrm{opt}} \tau^{(\mathrm{LO})}_{\mathrm{j}} \right)^2 \left( N^2 - 1 \right)}{12 N^2} , \quad (31)$$

where we used the relation $B_{\mathrm{opt}} \approx f_{\mathrm{FSR}}$.

In the next step we add the statistically independent jitter contributions from the ADC and the LO,

$$\mathrm{NSR}^{(\mathrm{jitter})} = \mathrm{NSR}^{(\mathrm{jitter})}_{\mathrm{ADC}} + \mathrm{NSR}^{(\mathrm{jitter})}_{\mathrm{LO}}$$
$$= \frac{\left( 2\pi B_{\mathrm{opt}} \right)^2 \left[ \tau^{(\mathrm{ADC})\,2}_{\mathrm{j}} + \tau^{(\mathrm{LO})\,2}_{\mathrm{j}} \left( N^2 - 1 \right) \right]}{12 N^2} . \quad (32)$$

If, for simplicity, we additionally assume identical rms-jitter values for the ADC and the LO comb, $\tau_{\mathrm{j}} = \tau^{(\mathrm{ADC})}_{\mathrm{j}} = \tau^{(\mathrm{LO})}_{\mathrm{j}}$, the above relation simplifies to

$$\mathrm{NSR}^{(\mathrm{jitter})} = \frac{\left( 2\pi B_{\mathrm{opt}} \tau_{\mathrm{j}} \right)^2}{12} = \frac{\left[ 2\pi \left( B_{\mathrm{opt}}/2 \right) \tau_{\mathrm{j}} \right]^2}{3} . \quad (33)$$

The broadband test signal, Eq. (22), is then obtained by taking the reciprocal of the right-hand side of Eq. (32) and (33), respectively.

### Acknowledgments

We acknowledge Artem Kuzmin for helpful discussions and for support in the experimental implementation.

### Funding

This work was supported by the ERC Consolidator Grant TeraSHAPE (# 773248), by the EU H2020 project TeraSlice (# 863322), by the DFG projects PACE (# 403188360) and GOSPEL (# 403187440) within the Priority Programme SPP 2111 "Electronic-Photonic Integrated Systems for Ultrafast Signal Processing" and via the DFG Collaborative Research Center (CRC) HyPERION (SFB 1527), by the BMBF project Open6GHub (# 16KISK010), by the EU H2020 Marie Skłodowska-Curie Innovative Training Network MICROCOMB (# 812818), by the Alfried Krupp von Bohlen und Halbach Foundation, by the MaxPlanck School of Photonics (MPSP), and by the Karlsruhe School of Optics & Photonics (KSOP)

### References

1.  N. K. Fontaine, R. P. Scott, L. Zhou, F. M. Soares, J. P. Heritage, and S. J. B. Yoo, "Real-time full-field arbitrary optical waveform measurement," Nature Photonics **4**, 248–254 (2010), doi: 10.1038/nphoton.2010.28.

2.  N. K. Fontaine, T. Sakamoto, D. J. Geisler, R. P. Scott, B. Guan, and S. J. B. Yoo, "Coherent reception of 80 GBd QPSK using integrated spectral slice optical arbitrary waveform measurement," in Conference on Lasers and Electro-Optics 2012, paper CTh1H.1, doi: 10.1364/CLEO_SI.2012.CTh1H.1.

3.  N. K. Fontaine, G. Raybon, B. Guan, A. Adamiecki, P. J. Winzer, R. Ryf, A. Konczykowska, F. Jorge, J.-Y. Dupuy, L. L. Buhl, S. Chandrashekhar, R. Delbue, P. Pupalaikis, and A. Sureka, "228-GHz coherent receiver using digital optical bandwidth interleaving and reception of 214-GBd (856-Gb/s) PDM-QPSK," in European Conference and Exhibition on Optical Communication (2012), doi: 10.1364/ECEOC.2012.Th.3.A.1.

4.  R. Proietti, C. Qin, B. Guan, N. K. Fontaine, S. Feng, A. Castro, R. P. Scott, and S. J. B. Yoo, "Elastic optical networking by dynamic optical arbitrary waveform generation and measurement," J. Opt. Commun. Netw. **8**, A171 (2016), doi: 10.1364/JOCN.8.00A171.

5.  Dengyang Fang, Andrea Zazzi, Juliana Müller, Daniel Drayss, Christoph Füllner, Pablo Marin-Palomo, Alireza Tabatabaei Mashayekh, Arka Dipta Das, Maxim Weizel, Sergiy Gudyriev, Wolfgang Freude, Sebastian Randel, J. Christoph Scheytt, Jeremy Witzens, and Christian Koos, "Optical Arbitrary Waveform Measurement (OAWM) Using Silicon Photonic Slicing Filters," J. Lightwave Technol., JLT **40**, 1705–1717 (2022), doi: 10.1109/JLT.2021.3130764.

6.  D. Drayss, D. Fang, C. Füllner, G. Likhachev, T. Henauer, Y. Chen, H. Peng, P. Marin-Palomo, T. Zwick, W. Freude, T. J. Kippenberg, S. Randel, and C. Koos, "Slice-less optical arbitrary waveform measurement (OAWM) in a bandwidth of more than 600 GHz," in Optical Fiber Communications Conference and Exhibition (2022), doi: 10.1364/OFC.2022.M2I.1.

7.  D. Drayss, D. Fang, C. Füllner, G. Lihachev, T. Henauer, Y. Chen, H. Peng, P. Marin-Palomo, T. Zwick, W. Freude, T. J. Kippenberg, S. Randel, and C. Koos, "Non-sliced optical arbitrary waveform measurement (OAWM) using soliton microcombs," Optica, **10**, 888-896 (2023)., doi: 10.1364/OPTICA.484200

8.  Drayss, D. Fang, C. Füllner, G. Lihachev, T. Henauer, Y. Chen, H. Peng, P. Marin-Palomo, T. Zwick, W. Freude, T. J. Kippenberg, S. Randel, and C. Koos, (2023). Supplementary document for non-sliced optical arbitrary waveform measurement (OAWM) using soliton microcombs. Optica Publishing Group. Journal contribution. https://doi.org/10.6084/m9.figshare.23643564.v3

9.  D. Drayss, D. Fang, C. Füllner, A. Kuzmin, W. Freude, S. Randel, and C. Koos, "Slice-less optical arbitrary waveform measurement (OAWM) on a silicon photonic chip," in European Conference on Optical Communication (2022), paper We4E6.

10. A. Zazzi, J. Müller, S. Gudyriev, P. Marin-Palomo, D. Fang, J. C. Scheytt, C. Koos, and J. Witzens, "Fundamental limitations of spectrally-sliced optically enabled data converters arising from MLL timing jitter," Opt. Express, OE **28**, 18790–18813 (2020). doi: 10.1364/OE.382832

11. A. Zazzi, J. Müller, M. Weizel, J. Koch, D. Fang, A. Moscoso-Mártir, A. T. Mashayekh, A. D. Das, D. Drayß, F. Merget, F. X. Kärtner, S. Pachnicke, C. Koos, J. C. Scheytt, and J. Witzens, "Optically enabled ADCs and application to optical communications," IEEE Open Journal of the Solid-State Circuits Society, **1** (2021), doi: 10.1109/OJSSCS.2021.3110943

12. D. Fang, D. Drayss, G. Lihachev, P. Marin-Palomo, H. Peng, C. Füllner, A. Kuzmin, J. Liu, R. Wang, V. Snigirev, A. Lukashchuk, M. Zhang, P. Kharel, J. Witzens, C. Scheytt, W. Freude, S. Randel, T. J. Kippenberg, and C. Koos, "320 GHz analog-to-digital converter exploiting Kerr soliton combs and photonic-electronic spectral stitching," in European Conference on Optical Communication (2021), doi: 10.1109/ECOC52684.2021.9606090

13. C. Deakin and Z. Liu, "Frequency interleaving dual comb photonic ADC with 7 bits ENOB up to 40 GHz," CLEO (2022), paper STh5M.1, doi: 10.1364/CLEO_SI.2022.STh5M.1

14. N. K. Fontaine, R. P. Scott, and S. Yoo, "Dynamic optical arbitrary waveform generation and detection in InP photonic integrated circuits for Tb/s optical communications," Optics Communications **284**, 3693–3705 (2011). doi: 10.1016/j.optcom.2011.03.045

15. J. Müller, A. Zazzi, G. Vasudevan Rajeswari, A. Moscoso Martir, A. Tabatabaei Mashayekh, A. D. Das, F. Merget, and J. Witzens, "Optimized hourglass-shaped resonators for efficient thermal tuning of CROW filters with reduced crosstalk," IEEE 17th International

Conference on Group IV Photonics (GFP) (2021), doi: 10.1109/GFP51802.2021.9673904

16. Tektronix, "Techniques for extending real-time oscilloscope bandwidth", White Paper 55W-29371-3 (2015), http://www.tek.com/document/whitepaper/techniques-extending-real-time-oscilloscope-bandwidth.

17. X. Chen, X. Xie, I. Kim, G. Li, H. Zhang, and B. Zhou, "Coherent detection using optical time-domain sampling," IEEE Photon. Technol. Lett. 21, 286–288 (2009), doi: 10.1109/LPT.2008.2010868.

18. L. D. Coelho, O. Gaete, and N. Hanik, "An algorithm for global optimization of optical communication systems," AEU - Int. J. Electron. Commun. 63, 541–550 (2009), doi: 10.1016/j.aeue.2009.02.009

19. J. K. Fischer, F. Ludwig, L. Molle, C. Schmidt-Langhorst, C. C. Leonhardt, A. Matiss, and C. Schubert, "High-speed digital coherent receiver based on parallel optical sampling," Journal of Lightwave Technology 29, 378–385 (2011), doi: 10.1109/JLT.2010.2090132.

20. M. Blaicher, M. R. Billah, J. Kemal, T. Hoose, P. Marin-Palomo, A. Hofmann, Y. Kutuvantavida, C. Kieninger, P.-I. Dietrich, M. Lauermann, S. Wolf, U. Troppenz, M. Moehrle, F. Merget, S. Skacel, J. Witzens, S. Randel, W. Freude, and C. Koos, "Hybrid multi-chip assembly of optical communication engines by in situ 3D nano-lithography," Light: Science & Applications 9, 71 (2020), doi: 10.1038/s41377-020-0272-5.

21. R. Schmogrow, B. Nebendahl, M. Winter, A. Josten, D. Hillerkuss, S. Koenig, J. Meyer, M. Dreschmann, M. Huebner, C. Koos, J. Becker, W. Freude, and J. Leuthold, "Error Vector Magnitude as a Performance Measure for Advanced Modulation Formats," IEEE Photonics Technology Letters 24, 61–63 (2012), doi: 10.1109/LPT.2011.2172405.

22. Keysight Technologies, M8135A & M8136A Universal Signal Processing Architecture (USPA) (data sheet), https://www.analog.com/media/en/technical-documentation/data-sheets/ad9209.pdf, (accessed January 2023).

23. Analog Devices, 12-Bit, 6 GSPS/10.25 GSPS, JESD204B, RF Analog-to-Digital Converter AD9213 (data sheet), https://www.analog.com/media/en/technical-documentation/data-sheets/AD9213.pdf (accessed January 2023).

24. Analog Devices, 12-Bit, 4 GSPS, JESD204B/JESD204C Quad ADC AD9209 (data sheet), https://www.analog.com/media/en/technical-documentation/data-sheets/ad9209.pdf, (accessed January 2023).

25. Texas Instruments, ADC12DJ5200-EP 10.4-GSPS Single-Channel or 5.2-GSPS Dual-Channel, 12-bit, RF-Sampling Analog-to-Digital Converter (ADC) (data sheet), https://www.ti.com/lit/ds/symlink/adc12dj5200-ep.pdf, (accessed January 2023).

26. Texas Instruments, ADC12DJ4000RF 8-GSPS Single-Channel or 4-GSPS Dual-Channel, 12-bit, RF-Sampling Analog-to-Digital Converter (data sheet), https://www.ti.com/lit/ds/symlink/adc12dj4000rf.pdf, (accessed January 2023).

27. Texas Instruments, ADC32RF5x Dual Channel 14-bit 2.6 to 3-GSPS RF Sampling Data Converter (data sheet), https://www.ti.com/lit/ds/symlink/adc32rf55.pdf (accessed January 2023).

28. Keysight Technologies, Datasheet: Infiniium UXR-Series Oscilloscopes, https://www.keysight.com/de/de/assets/7018-06242/data-sheets/5992-3132.pdf, (accessed January 2023)

29. Keysight Technologies, Infiniium V-Series Oscilloscopes (data sheet), https://www.keysight.com/de/de/assets/7018-04693/data-sheets/5992-0425.pdf (accessed January 2023).

30. Keysight Technologies, Infiniium S-Series (data sheet), https://www.keysight.com/de/de/assets/7018-04261/data-sheets/5991-3904.pdf (accessed January 2023).

31. Teledyne LeCroy, WaveMaster 8 Zi-B (data sheet) https://cdn.teledynelecroy.com/files/pdf/wavemaster-8zi-b-datasheet.pdf (accessed January 2023).

32. Teledyne LeCroy, WavePro HD (data sheet), https://cdn.teledynelecroy.com/files/pdf/waveprohd-datasheet.pdf (accessed January 2023).

33. Teledyne LeCroy, WaveRunner 8000HD (data sheet), https://cdn.teledynelecroy.com/files/pdf/waverunner-8000hd-datasheet.pdf, (accessed January 2023).

34. Teltronix, Serie MSO/DPO70000 (data sheet), https://download.tek.com/datasheet/DPO-DSA-MSO70000-DataSheet-GE-GE-55G-23446-36.pdf, (accessed January 2023).

35. R. H. Walden, "Analog-to-digital converter survey and analysis," IEEE J. Sel. Areas Commun. 17, 539–550 (1999).

36. R. H. Walden, "Analog-to-Digital conversion in the early twenty-first century," in Wiley Encyclopedia of Computer Science and Engineering (John Wiley & Sons, Ltd, 2008).

37. B. Murmann, "The Race for the Extra Decibel: A Brief Review of Current ADC Performance Trajectories," IEEE Solid-State Circuits Mag. 7, 58–66 (2015), doi: 10.1109/MSSC.2015.2442393

38. C. Pearson, "High Speed Analog to Digital Converter Basics," Application Report Texas Instruments (2011) http://www.ti.com/lit/an/slaa510/slaa510.pdf (accessed January 2023).

39. Texas Instruments "High-speed data converter fundamentals", Online Course https://www.ti.com/video/5529003238001#transcript-tab (accessed July 2023).

40. IEEE Instrumentation & Measurement Society "IEEE Standard for Terminology and Test Methods for Analog-to-Digital Converters", , IEEE Std 1241™ 2010, https://ieeexplore.ieee.org/stamp/stamp.jsp?arnumber=5692956 (accessed March 2023).

41. W. Kester, "Aperture time, aperture jitter, aperture delay time - removing the confusion," Analog Devices, MT-007, Tutorial, https://www.analog.com/media/en/training-seminars/tutorials/MT-007.pdf. (accessed January 2023)

42. Teledyne LeCroy, LabMaster 10 Zi-A High Bandwidth Modular Oscilloscopes (data sheet), https://cdn.teledynelecroy.com/files/pdf/labmaster-10zi-a-datasheet.pdf, (accessed January 2023)

43. A. Khilo, S. J. Spector, M. E. Grein, A. H. Nejadmalayeri, C. W. Holzwarth, M. Y. Sander, M. S. Dahlem, M. Y. Peng, M. W. Geis, N. A. DiLello, J. U. Yoon, A. Motamedi, J. S. Orcutt, J. P. Wang, C. M. Sorace-Agaskar, M. A. Popović, J. Sun, G.-R. Zhou, H. Byun, J. Chen, J. L. Hoyt, H. I. Smith, R. J. Ram, M. Perrott, T. M. Lyszczarz, E. P. Ippen, and F. X. Kärtner, "Photonic ADC overcoming the bottleneck of electronic jitter," Opt. Express 20, 4454–4469 (2012), doi: 10.1364/OE.20.004454

44. D. Fang, H. Peng, D. Drayss, Y. Chen, C. Füllner, A. Sherifaj, G. Lihachev, W. Freude, S. Randel, T. J. Kippenberg, T. Zwick, C. Koos, "Optical Arbitrary Waveform Generation (OAWG) Based on RF Injection-Locked Kerr Soliton Combs," ECOC 2023, Paper Th.B.2.6

45. Y. Chen, H. Peng, D. Fang, J. Dittmer, G. Lihachev, A. Voloshin, S. T. Skacel, M. Lauermann, A. Tessmann, S. Wagner, S. Bhave, I. Kallfass, T. Zwick, W. Freude, S. Randel, T. J. Kippenberg, C. Koos, "Self-Injection-Locked Kerr Soliton Microcombs With Photonic Wire Bonds For Use in Terahertz Communications", CLEO 2023, Paper STh3J.1

46. Arwa Hassan Beshr, "Study of ASE noise power, noise figure and quantum conversion efficiency for wide-band EDFA," Optik 126, 3492–3495 (2015), doi: 10.1016/j.ijleo.2015.08.225.

47. Y. Painchaud, M. Poulin, M. Morin, and M. Têtu, "Performance of balanced detection in a coherent receiver," Opt. Express, 17, 3659–3672 (2009), doi: https://doi.org/10.1364/oe.17.003659

48. W. Freude, "Optical Transmitters and Receivers", Lecture Notes (2022) pp.160-161

49. L. Lundberg, M. Mazur, A. Fülöp, V. Torres-Company, M. Karlsson, and P. A. Andrekson, "Phase Correlation Between Lines of Electro-Optical Frequency Combs," in Conference on Lasers & Electro-Optics (2023), paper JW2A.149. doi: 10.1364/CLEO_AT.2018.JW2A.149